\input{psfig.sty}
\documentclass{aastex}
\usepackage{spr-astr-addons}

\RequirePackage{color}

\begin{document}

\title{On the dark matter haloes inner structure and galaxy morphology}
\shorttitle{On the dark matter haloes inner structure and galaxy morphology}
\shortauthors{Del Popolo}

\author{A. Del Popolo\altaffilmark{1,2,3}} 

\altaffiltext{1}{Dipartimento di Fisica e Astronomia, University Of Catania, \\
Viale Andrea Doria 6, 95125 Catania, Italy}
\altaffiltext{2}{INFN sezione di Catania, \\ Via S. Sofia 64, I-95123 Catania, Italy}
\altaffiltext{3}{International Institute of Physics, Universidade Federal do Rio Grande do Norte, \\
59012-970 Natal, Brazil}

\begin{abstract}
In the present paper, we {extend the study of Del Popolo (2010) 
to determine the slope of the inner density profile of galaxy haloes 
with different morphologies.
}

We study how galaxy morphology changes the relation between the inner slope of the galaxy halo density profile, $\alpha$, and the stellar mass, $M_{*}$, or rotation velocity $V_{\rm rot}$. For this, we use the model of Del Popolo (2009) in combination with observed data from the Romanowsky \& Fall (2012) sample of elliptical and spiral galaxies, the Local Group sample compiled by McConnachie (2012), and the simulation results by Cloet-Osselaer et al. (2014). 

We find that the slope $\alpha$ flattens monotonically, from $\alpha \simeq -1 $ at $V_{\rm rot} \simeq 250$ km/s, to $\alpha \simeq 0 $. After $V_{\rm rot}\simeq 25$ km/s the slope starts to steepen.
The steepening happens in the mass range dominated by non-rotationally supported galaxies (e.g., dSphs) 
and depends on the level of offset in the angular momentum of rotationally and non-rotationally dominated galaxies.
The steepening is a consequence of the decrease in baryons content, and angular momentum in spheroidal dwarf galaxies.

We finally compare our result to the SPH simulations of Di Cintio. 
Our result is in qualitatively agreement with their simulations, with the main difference that the inner slope $\alpha$ at small stellar masses ($M_* \lesssim10^{8} M_{\odot}$) is flatter than that in their simulations. 

As a result, the claim that finding a core in dwarf galaxies with masses slightly smaller than $\simeq 10^6 M_{\odot}$, (as in the Di Cintio, or Governato, supernovae feedback mechanism) would be a problem for the $\Lambda$CDM model must be probably revised. 
\end{abstract}



\section{Introduction}


The $\Lambda$CDM model, frequently referred to as the standard model of Big Bang cosmology, due to its success in describing 
several properties of the Universe at large scales (Spergel et al. 2003, Komatsu et al. 2011; Del Popolo 2007; Del Popolo 2013, 2014a), has been shown to have some limits at smaller scales, namely the scales of galaxies (e.g., Moore 1994; Boylan-Kolchin, Bullok, and Kaplinghat 2011, 2012; Del Popolo \& Cardone 2012), and the central parts of galaxy clusters (Del Popolo \& Gambera 2000; Del Popolo 2002; Newman et al. 2013a,b)\footnote{In reality, the paradigm also shows other problems  (e.g., the ``cosmic coincidence problem" and the cosmological constant problem (Weinberg 1989; Astashenok, \& Del Popolo 2012).}.
 
The most often reported problems of the $\Lambda$CDM model are: the Cusp/Core problem (Moore 1994; Flores \& Primack 1994), the missing satellites problem (Klypin et al. 1999; Moore et al. 1999), and the Too-Big-To-fail (TBTF) problem (Boylan-Kolchin, Bullock, and Kaplinghat 2011, 2012).  

The quoted problems are interconnected and a unified solution can be obtained taking into account the role of baryons in the central parts of haloes (Zolotov et al. 2012; Del Popolo et al. 2014; Del Popolo \& Le Delliou 2014). 

In the present paper, we are mainly concentrated on some features of the Cusp-Core problem, discussed in the following. The {halo} density profiles of dwarf galaxies and LSBs are on average nearly flat (Burkert 1995; de Blok, Bosma, \& McGauch 2003; Swaters et al. 2003; Oh et al. 2011; Oh et al. 2010, 2011; Kuzio de 
Naray \& Kaufmann 2011; 
Cardone \& Del Popolo 2012), 
while simulations of the haloes density profiles in the $\Lambda$CDM model are cuspy. The inner density profile is $\rho \propto r^{\alpha}$ with $\alpha=-1$  
for the Navarro, Frenk \& White (NFW) (1996, 1997) profile, and an Einasto profile with $\alpha  \simeq -0.8$ at 100 pc for more recent simulations (e.g., Stadel et el. 2009, Navarro et al. 2010). Lensing and kinematics constraints in the brightest cluster galaxies (BCGs) obtained in several studies (Sand et al. 2002, 2004, Newman et al. 2013a,b), highlighted that even in some clusters the DM profile is flatter than a cuspy NFW profile.
  
Even if there is agreement on the previous discrepancy, it is clear that the inner profile of several dwarf galaxies or LSBs are not always flat (e.g., Simon et al. 2005; Oman et al. 2015), and a mass dependency has been noticed in the THINGS sample by 
de Blok et al. (2008)\footnote{The sample studied in that paper showed that larger galaxies, $M_B<-19$, differently from smaller ones, have profiles equally well described by cored or cuspy profiles.}.  
This mass dependence has been for the first time predicted by Ricotti 2003\footnote{Jing \& Suto 2000 found different slopes with degrees of steepness decreasing from galaxies to cluster of galaxies, but this result is probably dependent on the resolution (e.g., Navarro et al. 2010).}. More recent studies showed that the halo inner slope depends on the halo mass and on the ratio $(M_{\rm gas}+M_{*})/M_{\rm total}$, where $M_*$ is the stellar mass (Del Popolo 2010 (DP10), Cardone et al. 2011; Del Popolo 2011 (DP11); 
Di Cintio et al. 2014 (DC14)). Moreover, recently, Oman et al. (2015) noticed that 
the rotation curves of some dwarfs agree well with simulations, while others do not.

%

We should also add that, even nowadays, different techniques (e.g., spherical Jeans equation, Schwarzschild modeling, multiple stellar populations techniques) applied to similar or to the same object have sometimes given different results (e.g., Strigari et al. 2010; Breddels \& Helmi 2013, found a cuspy profile in Fornax, while Walker \& Pe${\tilde n}$arrubia 2011; Amorisco \& Evans 2012; Battaglia et al. 2008; Agnello \& Evans 2012, found a core). 
Also for larger objects than the quoted MW satellites discrepancies are evident (e.g., in the case of NGC2976, Simon et al. (2003) finds $-0.17 < \alpha < -0.01$, Adams et al. (2012) finds $\alpha = -0.90 \pm 0.15$, and Adams et al (2014) finds $\alpha = -0.53 \pm 0.14$ if stars are tracers of the potential, or $\alpha = -0.30 \pm 0.18$ if gas is the tracer\footnote{In the case of Simon et al. (2005) the slope $\alpha$ is obtained fitting a $V_{\rm rot} \propto r^{(2-\alpha)/2}$ to the rotation curve. In the case of Adams et al. (2012) the slope $\alpha$ is obtained with a 5 parameter model, and using as a DM density profile a power-law. In the case of Adams et al. (2015) the slope $\alpha$ is obtained fitting the profile to a generalized NFW profile.}).

So, it is clear that the determination of the slope $\alpha$ is not trivial, and the problem is even more complicated at the mass scale of dwarf galaxies\footnote{Note that in several cases, even using nowadays kinematic maps, there is disagreement 
on the slopes of the inner haloes (Simon et al. 2005; Oh et al. 2011b; Adams et al. 2014).}. 
Oman et al. (2015) proposed to characterize the cusp-core problem in terms of inner mass deficit instead of slope $\alpha$.
However, the only study in the literature using a different way of measuring the slope is that of Walker \& Pen$\tilde{a}$rrubia (2011) who used the parameter $\Gamma \equiv \frac{d \log M}{d \log r}<3-\alpha$ to calculate the slope of Fornax and Sculptor.

The determination of the inner structure of dwarf galaxies is of paramount importance, since going to smaller galaxy masses, dark matter domination increases, and the gas content decreases. 
Due to the DM domination, one expects that
the density profile should become very close to that obtained in N-Body simulations, namely an Einasto profile (e.g., Navarro et al. 2010). 

In fact, if the gas content of dwarves is very poor, the two most well known astrophysical solutions to the Cusp-Core Problem, namely the flattening due to  
``dynamical friction from baryonic clumps" (DFBC) (El-Zant et al. 2001, 2004; Ma \& Boylan-Kolchin 2004; Nipoti et al. 2004; 
Romano-Diaz et al. 2008, 2009; Del Popolo 2009 (DP09); Cole et al. 2011; Inoue \& Saitoh 2011; Nipoti \& Binney 2015), and  
``supernovae feedback flattening" (SNFF) of the cusp (Navarro et al. 1996; 
Gelato \& Sommer-Larsen 1999; Read \& Gilmore 2005; Mashchenko et al. 2006, 2008; Governato et al. 2010), will not be able to solve the problem.
 
In other terms, the study of the density profiles in dwarf galaxies can return important information on the nature of DM. If future observations (e.g., GAIA, TMT), will observe predominantly cored profiles in dwarf galaxies with $M_* < 10^6 M_{\odot}$, the probability that DM is cold are scanty, at least in some of the SNFF model (e.g., DC14). 

Since the study of the inner structure of small mass dwarf galaxies is important to better understand the nature of DM, in the present paper we want to study how the $\alpha$-$M_*$ relation is modified, especially at small masses 
($M_* \lesssim 10^{7}$) by the morphology of dwarf galaxies dominating at those masses.

In DP10, we showed how the inner slope of rotationally dominated galaxies changes with mass, in the DFBC scheme. 
In that paper, we assumed that the galaxies were rotationally supported at all masses. In reality as shown by Fall (1983) (F83), Romanowsky \& Fall (2012) (RF12), Fall \& Romanowsky (2013) (F13), and Teklu et al. (2015) (T15), the stellar specific angular momentum (AM), $j_*$, of galaxies depends on the morphology. {In the plane $j_*-M_*$, spiral (discs) and  elliptical galaxies are distributed on two tracks whose logarithmic slopes are $\beta \simeq 0.6$ (see Fig. 34 of RF12).} Ellipticals have a value of $j_*$ $\simeq 3-5$ times smaller than those of spirals, if the K-band mass to light ratio is neglected, and $\simeq 7$ if it is taken into account (RF12). 
Correcting for the variations of the $M_*/L_K$ with the B-V color, FR13 found $\beta =0.6 \pm 0.1$ and an offset among Spirals and ellipticals 30\% 
larger\footnote{Note that a theoretical explanation of the spin difference between spirals and galaxies was tried in the Gaussian peak formalism 
framework, on the basis of the different values of the peak heights, $\nu=\delta(0)/\sigma$, forming spirals, and ellipticals. In the previous expression, $\delta(0)$ is the central peak overdensity, and $\sigma$ is the mass variance (see Hoffman 1986; Heavens \& Peacock 1988; Catelan \& Theuns 1996; Del Popolo \& Gambera 1996).}.
%
Similar results were found by T15, who emphasized the smooth transition between discs and spheroids\footnote{In the present paper, a pressure-dominated stellar system is indicated with the term ``spheroid". Spheroids may have rigid body rotation. As usual, the spheroidal component of a spiral galaxy is usually termed ``bulge". Galaxies having just a spheroidal component are termed ``ellipticals". For precision's sake, even some ellipticals have a disk-like component, similarly to lenticular galaxies. Ellipticals and lenticulars are indicated as ``early-type" galaxies.}.

%
%
%

It is then important to understand how morphology change the $\alpha$-$M_*$ ($\alpha$-$V_{\rm rot}$) relation found in DP10, especially in the dwarf galaxies domain, and compare with the results of SPH simulations (e.g., DC14).

{This study will be performed using a sample obtained from RF12, McConnachie et al. (2012)(MC12), and the Cloet-Osselaer et al. (2014) (CO14) 
sample. The Cloet-Osselaer's sample is also constituted by the RF12, and some of the galaxies in the MC12 sample\footnote{We did not consider in the C015 sample the RF12 galaxies, and the ones of the MC12 sample common to the C015.}.

The sample contains dE galaxies (De Rijcke et al. 2005 (DR05) and van Zee, Barton \& Skillman 2004 (VZ04)), and dIrr and dSph galaxies (Worthey et al. 2004, 
De Rijcke et al. 2006, McConnachie \& Irwin 2006, Leaman et al. 2012, Kirby, Cohen \& Bellazzini 2012, Kirby et al. 2014,   and Hidalgo et al. 2013), plus C015 simulated galaxies.
The choice of a peculiar sample usually implies that the result could be sample dependent. However, our sample contains different morphologies, from Spirals to Ellipticals (in the case of normal galaxies), and dE, dIrr, dSph galaxies, isolated and satellites of large galaxies.
%
}

The paper is organized as follows. In Section 2, we describe what the theory developed in RF12 and FR13 predicts concerning the $j_*$-$M_*$ dependence on the galaxies morphology, and how $j_*$ is calculated.

In Section 3, we apply the RF12 theoretical models of Section 2 to find $j_*$ in the mass range $ 10^3 \lesssim M_* \lesssim 10^{12} M_{\odot}$, for the previous quoted samples.

%
%
%
In Section 4, we study how the $\alpha$-$M_*$, and $\alpha$-$V_{\rm rot}$ relations depend on morphology. Section 5 summarizes the results and draws conclusions.

\section{Morphology dependence of the $j_*$-$M_*$ relation}

Galaxies classification has been based, especially in the past, on morphology (e.g., Hubble classification and  similar (see Cappellari et al. 2011)), 
or color criteria. More recently the classification has been based on physical quantities as the AM L, connected to the galaxy rotation velocity, mass M, and energy E, related to their linear size. F83 introduced the so-called $j_*$-$M_*$ diagram, being $j_*=L_*/M_*$ the stellar specific angular momentum (SSAM), and $M_*$ the stellar mass. The diagram displays on the axes two independent variables, which are also conserved quantities. In the quoted diagram, as noticed by F83 and later by several other authors (Takase \& Kinoshita 1967; Heidemann 1969; Freeman 1970; Nordsieck 1973; RF12; FR13; T15), spirals are distributed on a tight band with $j_* \propto M_*^\beta$ ($\beta=0.6 \pm 0.1$, according to FR13). This observational result can be theoretically explained assuming that baryons constituting the galaxies and DM have the same $j$. The following dissipational collapse decouples baryons from DM and gives rise to star-formation (e.g., Ryden \& Gunn 1987). Assuming $j$ conservation, one can reproduce the slope 
{and} {zero point} of the $j_*$-$M_*$ relation. In the quoted diagram, ellipticals are distributed into a lower band parallel to that of spirals and more scattered (Bertola
\& Capaccioli 1975). Even if this behavior is not completely clear, the two band distribution is explained as a different capability of spirals and ellipticals to retain their initial $j$. According to RF12, the reasons producing this offset, and the trend of ellipticals can be explained in terms of multiple mergers, stripping, and outflows. The Hubble sequence, as proposed by F83, would be a variation in $j_*$ for a given $M_*$. 
If a galaxy is decomposed into bulge and disk component, the quoted $j_*$-$M_*$ parameter space originates a unique bulge fraction $B/T$, $M_*$ space. 
For a given $M_*$, one moves from the pure discs line ($B/T=0$) to the pure bulges ($B/T=1$), and this $B/T$ sequence should reflect the $j_*$-$M_*$ trend (RF12).  


{
RF12 calculated the specific angular momentum (SAM) with two different methods: 1) an exact one, based on the knowledge of the surface density $\Sigma(x)$, and the rotation velocity, and 2) an estimation, that they showed to give a very good evaluation of the SAM, $j_t$. }

In method 1, 
the SAM can be calculated as 
\begin{equation}
j_{\rm t}=C_{\rm i} j_{\rm p},
\label{eq:jt}
\end{equation}
where
\begin{equation}
j_{\rm p}= \frac{\int 
V_{\rm rot}(x) \Sigma(x)x^2 dx}{\int \Sigma(x)x dx},
\label{eq:jp}
\end{equation}
and where $\Sigma(x)$ is the surface density profile along the semimajor axis, $x$, and 
$V_{\rm rot}$ the rotational velocity along $x$.
The accuracy of the results of Eq. (\ref{eq:jp}) depends on the accuracy with which we know $\Sigma(x)$, and 
$V_{\rm rot}$. 

The projection term $C_{\rm i}$ is $1/\sin i$ for thin discs, being $i$, the inclination angle relative to the line of sight, and in general for spheroids we get
\begin{equation}
C_{\rm i} \simeq \frac{0.99+0.14i}{\sin i},
\label{eq:proj}
\end{equation}
(Eq. A29 in RF12). 

Using the previous formula, and assuming that $\Sigma(R) \propto \exp(-R/R_d)$, and $V_{\rm rot}= v_{\rm c}$, in the case of a thin disk one gets:
\begin{equation}
j_{\rm t}= 2 v_{\rm c} R_{\rm d}=1.19 v_{\rm c} R_{\rm e},
\label{eq:jtt}
\end{equation}
being $R_{\rm e}$ the half light radius ($R_{\rm e}=1.68 R_{\rm d}$). 

{
The circular velocity, $v_c$, present in the previous equations, must be evaluated at a given distance. For the reasons described in RF12, $v_c$ is evaluated at 
at radius $x_s \simeq 2 a_e$, being $a_e$ the effective radius along the semimajor axis. This velocity is indicated as 
$v_s$. Taking into account the inclination $i$, 
$v_s=V_{\rm rot} \sin{i}$, we have 
}
\begin{equation}
j_{\rm t}=2 V_{\rm rot} R_d=2 v_{\rm s} R_{\rm d}/\sin(i).
\end{equation}
In the case of spheroids, assuming a Sersic profile, $\Sigma(R) \propto \exp(-b_n(R/R_e)^{1/n})$, where $b_n$ is a function of the Sersic index $n$ (see Marquez et al. 2000), and in the case of a spheroid with a de Vaucouleurs profile ($n=4$), with {constant intrinsic velocity} $v_{\rm t}$, the AM is given by (Zasov 1985; RF12, Eq. A28)
\begin{equation}
j_{\rm t}=2.29 v_{\rm t} R_{\rm e}.
\label{eq:jttt}
\end{equation}
For a generic $n$, one has to  use Eq. (\ref{eq:jt}) and Eq. (\ref{eq:jp}) with Eq. (\ref{eq:proj}).

The method 2, is a very good estimator of $j_t$. The projected part of $j_t$, dubbed 
$j_{\rm p}$, is given by (RF12)
\begin{equation}
\tilde{j}_{\rm p}= k_n v_s R_e,
\label{eq:esti}
\end{equation}
where
\begin{equation}
k_n \simeq 1.15+0.029 n+0.062 n^2,
\label{eq:estim}
\end{equation}
which for $n=1$ and $n=4$ gives values close to 1.19 and 2.29, given in Eq. (\ref{eq:jtt}) and Eq. (\ref{eq:jttt}) (RF12).
$j_t$ is obtained from Eq. (\ref{eq:estim}), recalling that $j_t= C_i j_p$.

{ 
Looking at RF12, Table 3, a simple calculation shows that the value of $j_p$, obtained with the method 1, is in very good agreement to that obtained with the estimator.
}

~\\

{
The value of the specific AM of RF12 sample was obtained through this estimator (see Tab. 4, 5 in RF12).


More in detail, in the case of pure discs, Eq. (\ref{eq:jtt}) was used. Note that this is equivalent to calculate $\tilde{j}_{\rm p}$, for $n=1$ (exponential disk), which gives $\tilde{j}_{\rm p} \simeq 2 v_s R_d$. Then $j_t= \tilde{j}_{\rm p}/\sin(i)=2 v_s R_d/\sin(i)=2 v_c R_d$. 
%
In systems with large bulges, the contribution to $j_t$ coming from the bulge and disk was separately calculated and then combined (see their Section 4.1). 
{ 
If $j_b$ and $j_d$ are the SAM of the bulge and the disk, respectively, { 
and a bulge stellar mass fraction quantified as $f_b$,
the total SAM is given by
\begin{equation}
j=f_b j_b+(1-f_b) j_d
\end{equation}
The bulge-to-total luminosity ratio $B/T$ is used as a proxy of $f_b$ (RF12).
}

Both $j_b$ and $j_d$ were calculated with the quoted estimator,
$\tilde{j}_{\rm p}$,  Eq. (\ref{eq:esti}). Knowing the inclination $i$ (assumed equal for the bulges and the discs) $j_t$ was obtained. }
%
%
In the case of bulges, $j_*$ was estimated through $\tilde{j}_{\rm p}$, and $v_s$ was obtained through Eq. (8) of RF12. More precisely, it is given by 
\begin{equation}
v_s=(v/\sigma)^* \sigma_0 \left( \frac{\epsilon}{1-\epsilon} \right)^{1/2}
\label{eq:vel}
\end{equation}
assuming $(v/\sigma)^*=0.7$, where $(v/\sigma)^*$ is a parameter indicating the relative importance of pressure and rotation, $\sigma_o$ is the central velocity dispersion, $\epsilon=1-b/a$ is the ellipticity, and $b$, and $a$ the semi-minor and semi-major axes, respectively. 
In the case of elliptical galaxies, it was used again the estimator $\tilde{j}_{\rm p}$. $j_t$ was obtained using Eq. (\ref{eq:jt}). The projection term $C_i$ is an average value, equal to 1.21 for the lenticulars and 1.65 for the ellipticals (RF12).
 
} 

%
%

{ 
%

The value of $j_*$ for McConnachie's sample was calculated in a similar way (see Section 3.1). 

}

~\\

{
The angular momenta obtained were connected, by RF12, to the tidal torque theory (TTT) predictions modified to take into account the transformation of gas into stars, and the different quantity of AM lost by baryons for different morphologies. They showed that in the $j_*-M_*$ plane galaxies of different morphologies are distributed on parallel lines, similarly to their Fig. 2. A summary of this model can be found in Appendix C.}

\section{Distribution of spiral, elliptical, and  dwarf galaxies in the $j_*$-$M_*$ plane}


In the previous section, we saw how to calculate $j_*$ for the RF12 sample, and that the AM of discs and spheroids distributes { on two parallel bands}\footnote{Note that considering not only discs but spirals of different morphologies, the best fit gives rise to lines slightly diverging (see Fig. 14a of RF12).}. However, the previous studies (RF12, FR13) concerns normal galaxies with masses down to $10^{8.4} M_{\odot}$. In the case of dwarf galaxies (dEs, dIrrs, dSphs) there are not many studies. Cloet-Osselaer et al. (2014) (see their Fig. 13) compared simulations and observations of dwarf galaxies starting from a stellar mass $M_*=10^{5} M_{\odot}$ to giant galaxies ($10^{12} M_{\odot}$). {An important result they got is 
that the results of the previous section, namely RF12 and FR13, extends to dwarf galaxies (see their Fig. 13) (see next subsection). }

%

%
%

%
%


\subsection{$j_*$-$M_*$ relation for dwarves, spirals, and ellipticals}

\begin{figure*}
\centering
\includegraphics[width=160mm]{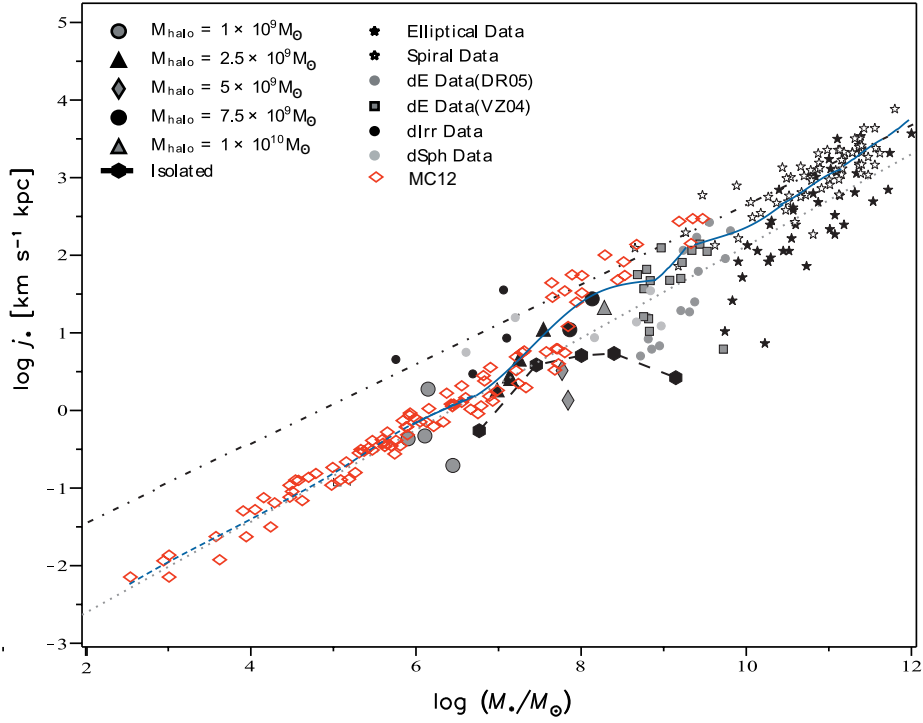} 
\caption[]{$j_*$-$M_*$ relation. The stars represent the RF12 sample, the red squares that of MC12. The remaining symbols represents data coming from the Cloet-Osselaer sample, constituted (apart the MC12 galaxies already considered) by  dE galaxies (De Rijcke et al. (2005)(DR05) and van Zee, Barton \& Skillman (2004)(VZ04)), and dIrr and dSph galaxies (Worthey et al. 2004, 
De Rijcke et al. 2006, McConnachie \& Irwin 2006, Leaman et al. 2012, Kirby, Cohen \& Bellazzini 2012, Kirby et al. 2014,   and Hidalgo et al. 2013), plus their simulated galaxies. The dot-dashed, and the dotted line are the fits for the spirals and ellipticals, respectively. The solid line is the average $j_*$ calculated as discussed in the text.
}
\end{figure*}

In this section, we find the distribution of galaxies in the $j_*$-$M_*$ plane in the mass range $10^3-10^{12} M_{\odot}$.
For the range $10^{8.4}-10^{12} M_{\odot}$, we used the sample compiled by RF12 (Table 4, Table 5).
%
%

{
The specific angular momenta $j_*$ for the sample of RF12 (spirals, ellipticals, and intermediate configurations) were obtained as described in Section 3.
}

%
%
%
%

%

{
In the case of the McConnachie's sample, we used a similar method to that of RF12. Namely, we used 
Eq. (\ref{eq:jtt}) for the few spirals in MC12 sample, 
as MC12, and for the remaining galaxies we used the 
$\tilde{j}_{\rm p}$ estimator. To calculate the quoted estimator, we used the values of $a_e$, $\epsilon$, and the dispersion velocities given by MC12. $v_s$ was determined through Eq. (\ref{eq:vel}), with a different $(v/\sigma)^*$.

{
Bender \& Nieto (1990) (Table 3) and Mateo 1998 (Table 4) give the values of $(v/\sigma)$ for 20 galaxies. 
For the remaining galaxies, following Ferguson \& Binggeli (1994), Bender \& Nieto (1990) (Fig. 2), and CO14 (Fig. 15, and Tab. 3), we assumed a median value of $(v/\sigma)^*=0.5$. The previous choices for $(v/\sigma)^*$, only for a part of the MC12 sample, are following the same logic of RF12, and are subject to the same limits of that paper. For the galaxies for which they did not know the bulge rotations, they estimated it, as already discussed, through their Eq. 8 by using the $(v/\sigma)^*$ estimated from the $(v/\sigma)$-$\epsilon$ relation in their Fig. 10. We did the same using Fig. 15 of CO14, obtaining $(v/\sigma)^*=0.5 \pm 0.3$.

We followed RF12 also for what concerns the Sersic index $n$ determination. For the early type galaxies of their sample for which they did not know $n$, RF12 used a relation between $n$ and $M_*$ obtained using Fig. 10 of Graham \& Guzman (2003), and then converted the blue magnitude into $M_*$
(RF12, Eq. C1), for the mass range of their sample. 
%
%

We used Eq. (12) in Graham et al. (2006) (extension of Graham \& Guzman 2003), giving the relation $n-M_*$ till $M_* \simeq 10^7 M_{\odot}$, with a scatter
%
%
of $\sigma \simeq 1$, similar to the variations in the Sersic fits to galaxy of this kind. The uncertainty in the Sersic index translates into a 0.1 dex uncertainty in $j$ (Eq. (\ref{eq:esti})).

For smaller masses, we assumed $n=0.5 \pm 0.3$, according to Schroyen et al. (2011) (Fig. 16) simulations\footnote{A similar value is obtained in the CO14 simulations (Fig. 5.)}.

}
%
%
%
%
~\\


In the case of Cloet-Osselaer's sample, we used their evaluation, which followed the same method we used, namely that of RF12 (private communication).

~\\

{In Fig. 1, we plot the $j_*$-$M_*$ relation for the entire sample. The dot-dashed and the dotted line are the fits for the spirals, and ellipticals, respectively. The solid line is the average $j_*$ calculated as discussed in Section 4.2.
%
}


As noticed by FR13, in the case of the RF12, and FR13, the values of $j_*$ for ellipticals and spirals are the most robust ones, since in the case of morphologies intermediate between ellipticals and spirals, it is necessary to distribute mass and AM between bulges and discs of the ``composite" galaxies, so introducing uncertainties. 

{

The plot shows several interesting features: a) as already reported, discs and ellipticals are distributed on parallel lines
(recall the previous footnote: in the general case of spirals of different morphologies, the lines are not parallel). b) 
Isolated dwarf galaxies, coming from CO14 sample, distribute in the plane $j_*$-$M_*$ according to RF12 prediction (as also seen in Fig. 13 of CO14)\footnote{Almost all dwarf galaxies used by CO14 are isolated objects.}.
This is in agreement with Bender \& Nieto (1990) (see also Ferguson \& Binggeli 1994). Namely, the SAM of dwarf galaxies, $j_*$, fall onto the $j_*-M_*$ relation of 
ellipticals. c) The data coming from CO14 simulations, especially in the case of mergers,     
follow the trend of observed dwarf galaxies, which by converse follow the trendline of the elliptical galaxies (see last paragraph of Section 4.2 of CO14). 
%

%



}

%
%
%
%

\section{Dependence of the $\alpha$-$M_*$ relation from morphology}

As shown in several past papers of the same author (DP09; DP10; Del Popolo 2012a,b), {using the method described in the Appendix A of this paper}, the inner slope of structures (galaxies and clusters) depends on several physical parameters, $\alpha=f(M,h,j_r,\mu,F_b)$, where $M$ is the structure mass, $h$ is the DM ordered SAM, acquired by structures by tidal interactions with neighbors, $j_r$ is the random SAM, $\mu$ is the coefficient of dynamical friction (DF), and $F_b$ is the baryons content (gas, stars).

%

In several papers (DP09, DP12a,b, Williams et al. 2004, DC14, Polisenky \& Ricotti 2015) it was shown that more massive haloes have steeper profiles, and AM plays an important role in shaping them.

This result was recently checked in numerical simulations by Ricotti \& Polisenky (2015), who found results in agreement
with the previously quoted idea. 

The other factor that influences the profile structure is DF, which transfer energy and AM 
from baryons to dark matter. The larger is the coefficient of DF, $\mu$, the flatter is the profile (DP09, Del Popolo et al. 2014). 

{
Finally, for a fixed value of the mass (and the other parameters) the larger the content of baryons in a system the flatter is its profile (DP09; DP12a,b). Note with increasing mass,
from dwarf galaxies to giant galaxies, the baryon content increases, but its role in flattening the profile is overwhelmed by the steepening process produced by the mass increase (see Appendix A1 for a discussion on this issue.}

At this stage, we want to recall that the {model we use (sections A1-A5) is considering the interaction of the galaxies with their neighbors through tidal torques only. Environmental effects like tidal stripping, stirring, merging of the satellite with the host are taken into account as in  
Del Popolo \& Le Delliou (2014).
In the quoted paper, apart the model of cusp-core transformation (DP09, and Del Popolo \& Hiotelis 2014) (sections A1-A5) the interaction of satellites with environment is provided by a second semi-analytic model taking into account tidal stripping, tidal heating, dynamical friction. This model is summarized in  section A6. }

In the following, we show how the $\alpha$-$M_*$, or similarly the $\alpha$-$V_{\rm rot}$ relationship changes when we take into account the dependence of the AM of a structure from its morphology. 

We shall compare the results of our model, which is a DFBC, and which we will be indicated hereafter with ODFBC,  with the results of DC14 simulation.

%


\subsection{Two corrections to the DC14 $\alpha$-$V_{\rm rot}$ relation.}


As already discussed, in DP10 we calculated the dependence of the inner slope from the mass, in structures with masses ranging from dwarves to clusters of galaxies, studying how the structure changed in presence or absence of baryons. In structures containing DM and baryons, a strong dependence of the inner slope on mass, and redshift was found. The inner slope was shown to be an increasing monotonic function of mass and redshift. This leads to the conclusion that density profiles are no longer universal, when the structure contains baryons. The relation was calculated for non-rotational supported galaxies. In this section, we extend 
the slope-mass relation found in Fig. 2a of DP10 (solid line) to a {$\alpha-M_*$ relation dependent on morphology. We also calculated the $\alpha-V_{\rm rot}$ relation, using the rotational velocity given by the model described in Appendix A. Finally, we compared the results with those of DC14.}

%
%

%

{In DC14, the authors, using the code GASOLINE, showed that $\alpha$ depends on $M_*/M_{\rm halo}$. 
For $M_*/M_{\rm halo} \lesssim 0.01 \% $ the feedback energy is not enough to convert cusps into cores, and for larger 
$M_*/M_{\rm halo}$, the slope $|\alpha|$ tends to smaller values (less steep profile). The flattest profile is obtained for 
$M_*/M_{\rm halo} \simeq 0.5\%$. The slope starts to be larger for $M_*/M_{\rm halo} \gtrsim 0.5\%$, due to the fact that larger galaxies have more stars and deeper potential wells that tend to reduce the effects of feedback.

The relationship between $\alpha$ and $M_*/M_{\rm halo}$ was converted to a relation $\alpha-M_*$ using Moster et al. (2013) relation, and finally this was converted to a $\alpha$-$V_{\rm rot}$ relation by means of the Tully-Fisher relation (TFR) (Eq. (\ref{eq:duttt})). }


\begin{figure*}
\centering  \hspace{-1cm}
\includegraphics[width=169mm,height=70mm]{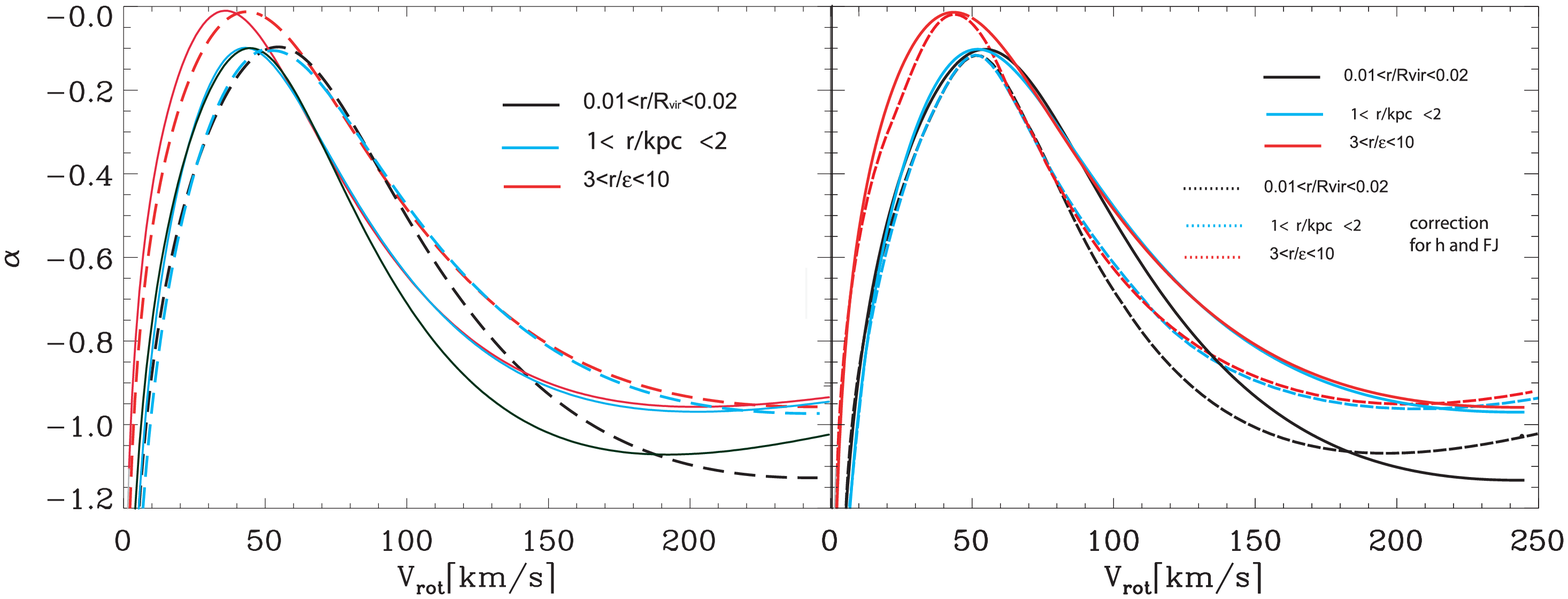}
\caption[]{The $\alpha$-$V_{\rm rot}$ relation. Panel a: the solid lines are those calculated by DC14, and the dashed ones, those corrected for the $h^{-2}$factor. The slope was calculated in the radial range a) 0.01 $< r/R_{\rm vir}<$ 0.02 (black lines). In DC14 simulations, this corresponds to 0.60 kpc $< r <$ 1.20 kpc, for the smallest mass, and to 1.30 kpc $< r <$ 2.65 kpc for the largest mass galaxy. b) $1 < r/kpc < 2$ (cyan lines). c) $3 < r/\epsilon < 10$ (red line), corresponding to 0.23 kpc $< r <$ 0.78 kpc for the smallest  DC14 halo, and 0.94 kpc $< r <$ 3.13 kpc for the largest. Panel b: 
the solid lines is the $\alpha$-$V_{\rm rot}$ relation of DC14, while the dotted ones are the correction to the previous relation correcting for the $h^{-2}$factor, and using the FJ relation for the non-rotating galaxies instead of the TFR, as done by DC14.
}
\end{figure*}



Before calculating the previous relations, and compare to DC14 results, we make two small corrections to Fig. 6 in DC14. 

Firstly, in order to obtain a $\alpha$-$V_{\rm rot}$ relation from the $\alpha$-$M_{*}$ one, DC14 used the TFR of Dutton (Eq. \ref{eq:duttt}), which contains a factor $h^{-2}$. This was forgotten by the authors when they made the conversion from $M_*$ to $V_{\rm rot}$ plotted in Fig. 6 of DC14. 

In Fig. 2 the results of Fig. 6 of DC14 are represented by solid (red, blue and black) lines, and the corrected curves are represented by 
dashed lines.
The radial range in which $\alpha$ was calculated is reported in the figure, for each curve. 


The quoted correction is plotted in Fig. 2, left panel, plotting $\alpha$ vs. $V_{\rm rot}$. 
%

In the plot, DC14 calculated the slope in the radial range $0.01 <r/R_{\rm vir}< 0.02$ (black lines), 
corresponding to the radial range 0.60 kpc $< r <$ 1.20 kpc, for the smallest mass in their simulation, and to 1.30 kpc $< r <$ 2.65 kpc for the largest mass galaxy. Similarly, the cyan lines correspond to $1<r/{\rm kpc}<2$, and the red lines to $3<r/\epsilon<10$, corresponding to the range 0.23 kpc $< r <$ 0.78 kpc for the smallest halo, and 0.94 kpc $<r <$ 3.13 kpc, for the largest halo.

%

The difference $\Delta \alpha$, after the correction, depends on the value of $V_{\rm rot}$ at which is measured. For example in the case of the red line (3 $<r/\epsilon <$ 10) the correction has a value $\Delta \alpha \simeq 70\%$, at $\simeq 25$ km/s, and $\Delta \alpha \simeq 40\%$ at 80 km/s. 

Apart the previous correction relative to the missing factor $h^{-2}$, another correction to Fig. 6 in DC14 should be made. 
In order to convert $M_*$ to $V_{\rm rot}$, DC14 used the TFR of Dutton et al. (2010),
\begin{equation}
\log_{10}{\frac{V_{2.2}}{\rm km/s}}=2.143+0.281 \left(\log_{10}{\frac{M_*}{10^{10} h^{-2} M_{\odot}} }\right)
\label{eq:duttt},
\end{equation}
where $V_{2.2}$ is the rotation velocity, $V_{\rm rot}$, at 2.2 $R_d$ (in the following, we indicate it as $V_{\rm rot}$).
{Despite the TFR of Dutton et al. (2010) is valid for $M_* > 10^{8.6} M_{\odot}$, DC14 assumed it to be valid for smaller masses\footnote{We know that in the case of dwarf galaxies the TFR is different from the one of normal galaxies.} and used it to extrapolate the $\alpha-M_*$ 
relation below $10^{8.6} M_{\odot}$. The dashed line in their Fig. 6 indicate this assumption. Apart the improper assumption, 
as already discussed, at small masses the mass function is dominated by non-rotational supported galaxies (e.g., dSphs), and for those objects 
one is not allowed to use the TFR. They should have used 
the Faber-Jackson relation (FJR) 
(Eq. (7) of Dutton et al. (2010))
\begin{equation}
\log \frac{\sigma(R_{50}}{\rm km/s})=2.054+0.286 \left(\log \frac{M_*}{\rm 10^{10} h^{-2} M_{\odot}}\right),
\label{eq:dutttt}
\end{equation}
where $R_{50}$ is the $r$-band half radius. The circular velocity at the projected half radius, $V_{50}$ can be obtained, according to Dutton et al. (2010), as $V_{50}=V_{\rm rot} (R_{50})=1.65 \sigma(R_{50})$.

The FJR of Dutton et al. (2010), similarly to their TFR, is valid till $10^{8.6} M_{\odot}$, and 
the FJR for dwarf galaxies is different from normal galaxies. However, we will use it following the DC14 assumption, since we want to compare our model's result concerning the inner slope of the density profile with theirs, shown in DC14, Fig. 6.}

In Fig. 2, right panel, we show how the $\alpha$-$V_{\rm rot}$ relation obtained by DC14 is changed by the two previous corrections ( a) missing $h^{-2}$in the conversion $M_*$-$V_{\rm rot}$); b) use of the FJR instead of the TFR, that was used for spheroidals by DC14).
Since the correction b) acts only at small velocities, $\Delta \alpha$ for $V_{\rm rot} \gtrsim 40$ km/s is the same to that calculated in the case of the correction $h^{-2}$ alone (Fig. 2, left panel), namely $\Delta \alpha \simeq 40\%$ at 80 km/s. 
At $\simeq 25$ km/s, $\Delta \alpha$ is 40\%, in the case of the red line. Note that $\Delta \alpha$, when taking into account both the corrections, at small velocities is smaller than that given when the two corrections are applied. This is due to the fact that while the first correction (missing $h^{-2}$) moves the curves to the left, the second correction (FJ correction) moves them to the right bringing them closer to the uncorrected DC14 lines.

\subsection{The $\alpha$-$M_*$ relation and morphology}



\begin{figure*}
\centering{}
\includegraphics[width=1.\hsize,]{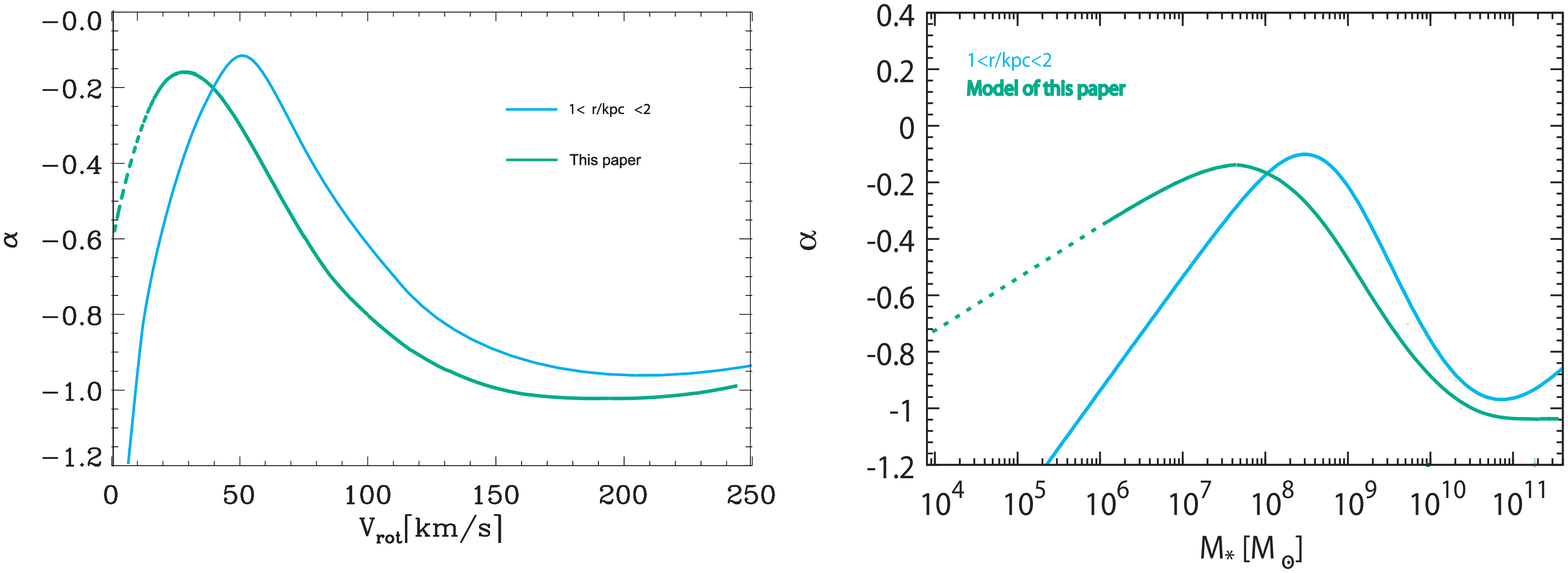} \hfill
\includegraphics[width=1.\hsize,]{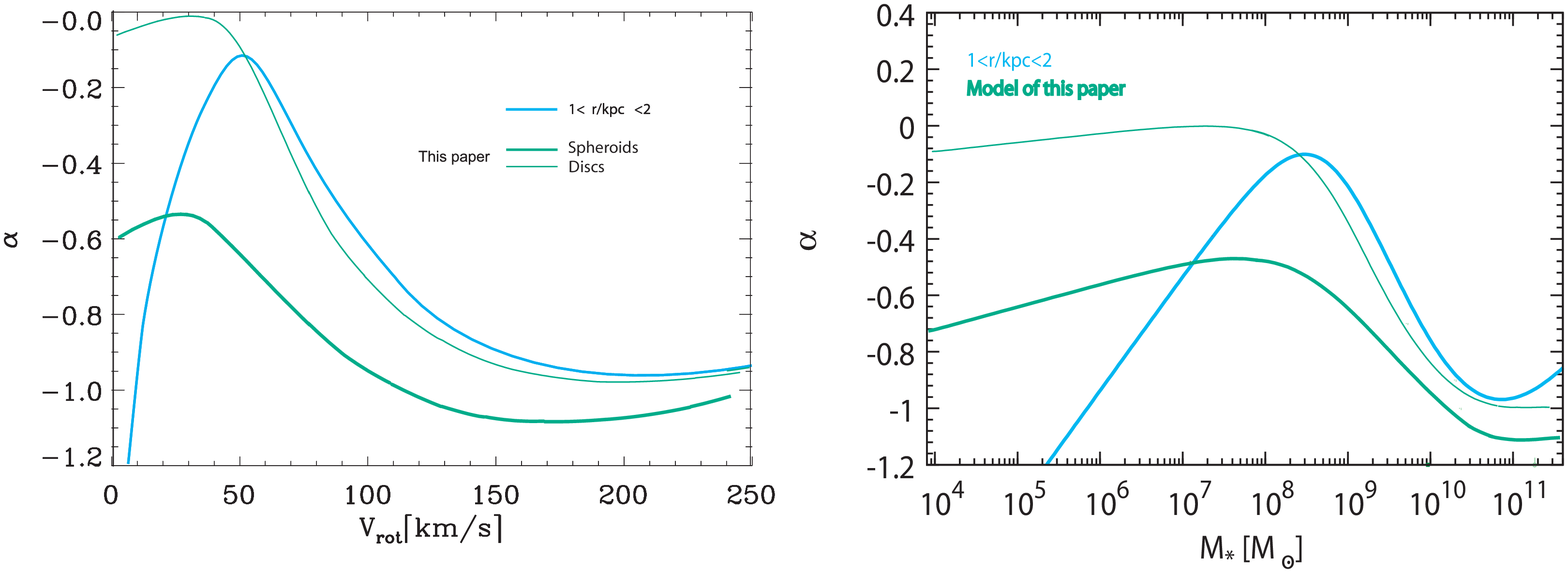} \hfill
\caption[]{The $\alpha$-$V_{\rm rot}$, $\alpha$-$M_{*}$ relations. Top left panel: the cyan line represents the $\alpha$-$V_{\rm rot}$ calculated by DC14, while the green line (counterpart of the DC14 cyan line) 
represents the $\alpha$-$V_{\rm rot}$ according to our model taking into account the difference in morphology of galaxies. 
Top right panel: 
as in the left panel, but for the $\alpha$-$M_*$ relation. The previous calculation is based on the average AM calculated for masses larger than $10^6 M_{\odot}$. The dashed line corresponds to the average AM for masses smaller than $10^6 M_{\odot}$.  Bottom left panel: same as top left panel, but now the thin and thick green lines represent the $\alpha$-$V_{\rm rot}$ relation for discs, and spheroids, respectively. Bottom right panel: as the left panel but for the $\alpha$-$M_*$ relation.
}
\end{figure*}

{
In order, to calculate the slope-mass relation, $\alpha-M_*$ (or $\alpha-V_{\rm rot}$), of our model and compare it 
with that of DC14, we employ our model (ODFBC), to ``simulate" the formation and evolution of galaxies having final stellar mass, $M_*$, and halo mass, $M_{\rm halo}$ similar to those of DC14. In order to reproduce the $j_*$-$M_*$ relation in Fig. 1, we simulated galaxies with the same stellar mass, $M_*$, and AM of those in Fig. 1. The stellar masses, for giant and dwarf galaxies are directly given by the model in Appendix A, and not calculated assuming a given value of the $M/L_{\rm V}$ ratio.}

For each galaxy, we calculated the inner slope of the density profile in the same DC14 radial bins.

{At this stage, we could calculate for each galaxy in the sample with a stellar SAM given in Fig. 1, the inner slope of the DM density profile. We would obtain then a scatter plot of the values of $\alpha$ in terms of $M_*$, or $V_{\rm rot}$. We prefer to follow a different path. We calculated an ``average value" of $j_*$, in the entire mass range of the sample. 

This can be done considering a bin in mass $M_*$ of variable width, chosen in order not to have too large jumps of $j_*$ from one bin to the other, and then calculating the average $j_*$.
This value is associated to the central value of the mass in the bin. In this way, we get an average value for $j_*$ in all the mass range of Fig. 1. This is shown by the solid line in Fig. 1.

{
The previous approach is strictly valid for a complete sample. For masses larger than $10^6 M_{\odot}$, the sample shown in Fig. 1, as described in the Introduction, is constituted by giant Spirals, Ellipticals and intermediate configurations, and dwarf galaxies. 
%
%

In the following, we assume that for masses larger than $10^6 M_{\odot}$ the sample is complete. We calculate the average $j_*$ as discussed and then the $\alpha$-$M_*$($V_{\rm rot}$) relation. At the same time, in order to consider the entire sample and avoid completeness issues,
we repeated the calculation for the two extreme sequences in the $j_*$-$M_*$ plane (discs, and spheroids).
}

Before moving forward, we want to recall that the $\alpha-V_{\rm rot}$ relation was obtained using the rotational velocity given by the model described in Appendix A.}


The results of the calculation of the $\alpha$-$V_{\rm rot}$, and $\alpha$-$M_*$ using our model is presented in Fig. 3. In this figure, the cyan line represents the DC14 result, while the green line (corresponding to the cyan one in DC14 (Fig.6)) are obtained with our model, as described previously. 
 
{ 
As previously reported, in the top left panel of Fig. 3, we plotted the $\alpha-V_{\rm rot}$ relation using the average AM calculated previously, assuming the completeness of the sample for masses $>10^6 M_{\odot}$.


}

The plot shows that at high velocity the slope $\alpha$ tends to -1, like the NFW model. In the case of the DC14 model, as previously described, this tendency is produced by the deepening of the potential well in normal galaxies which makes the feedback less effective. 

In our case this is due to the already described steepening of the slope with galaxy mass. 
%
Similarly to DC14, the stars content decreases, the potential well becomes shallower, and feedback is more efficient. 

In the velocity range 25-100 km/s the ODFBC predicts steeper profiles than the SNFF model, with a difference in slope $\Delta \alpha \simeq 0.15$. In the range 70-100 km/s, $\Delta \alpha \simeq 0.2$, and at $V_{\rm rot} \simeq 50$ km/s, $\Delta \alpha \simeq 0.3$.

At smaller velocities, $V_{\rm rot} \simeq 40-45$ km/s
the DC14 model predicts a maximum for the $\alpha$-$V_{\rm rot}$. 
At even smaller velocities the energy from SN feedback keeps decreasing and the slope steepens again, until it reaches $\alpha \simeq -1.2$.

In our model, the slope has a similar behavior: it reaches a maximum and then it steepens, due to the decrease of baryons content and AM when moving to the dSphs region.

{ 
More in detail, the plot shows that moving from fast rotators to slow rotators
%
%
the slope $\alpha$ becomes steeper\footnote{This is due to the fact that the density profile becomes steeper with decreasing AM.}.  
%

We see that at $V_{\rm rot} \lesssim 30$ km/s the slope monotonically decreases, for the reasons already discussed. 
%
%

}

Despite the similitude in behavior, 
the ODFBC model predicts 
slopes flatter than the SNFF, at small masses.

{
The $\alpha$-$V_{\rm rot}$ relation was calculated using the average AM obtained from Fig. 1 for masses larger than $10^6 M_{\odot}$. For smaller masses, the average AM is obtained only from the MC12 sample, which is mainly constituted by satellites of the LG, and then the sample is incomplete. For those masses (velocities), the relation is represented by a dashed line. 

}

%

The behavior at very small $M_*$ is better seen in the top right panel of Fig. 3, representing the same plots of the left panels but for $M_*$.



As in the top left panel, the green 
line is the prediction of the ODFBC.

The cyan line is given by\footnote{The equations $\alpha_{\rm cyan}$ have been provided by A. Di Cintio.}

\begin{eqnarray}
  \alpha_{\rm cyan}=0.167619-\log\left[\left(10^{X+2.14248}\right)^{-0.699049}+\right.\\
  \left.\left(10^{X+2.14248}\right)^{1.56202}\right]\;,
\end{eqnarray}
with $X$ being
\begin{equation}
 X=\log{\left[\frac{0.0702}
 {\frac{10^{6.7120}M_{\ast}^{-0.57912}}{4.6570}+\frac{10^{-2.9657}M_{\ast}^{0.25589}}{0.50674}}\right]}\;.
\end{equation}

{
While DC14 predicts cuspy profiles, namely $\alpha \simeq -1$ at $M_* \simeq 10^6 M_{\odot}$ (cyan line),
the ODFBC shows a much flatter profiles, $\alpha \simeq -0.4$, for the green line corresponding to cyan line of DC14.

%
}

The difference between the two models at small mass scales is due to the fact that in dSphs star formation efficiency is low. 
In the SNFF mechanism gas clumps must be converted into massive stars, then have to explode in SNs which will inject gas in the galaxy DM. In the ODFBC mechanism gas clumps can directly exchange energy and AM with DM before forming stars.  

{
In the bottom panels, we represent the same relations of the top ones but now the thin and thick green lines represent the $\alpha$-$V_{\rm rot}$, and $\alpha$-$M_*$ relation for discs and spheroids, respectively. As expected, in the case of discs, having larger AM than spheroids, the slope is flatter than in the case of spheroids, for all velocities, and masses. 
In the case of discs, the slope is steeper than DC14 simulations till masses $\simeq 50$ km/s, and then flatter. A similar behavior is shown by spheroids, when comparing with DC14 simulations. As previously noticed, spheroids show a slope steeper than the discs for all velocities and masses.
}

\section{Summary and discussion}

\subsection{Summary}

In the present paper, we studied how galactic morphology modifies the $\alpha$-$M_*$ relation in the ODFBC model, and compared the results with the SPH simulations of the quoted relation performed by DC14. As a first step, we discussed how the $j_*$-$M_*$ relation changes with morphology. 

As already known, spiral galaxies have a larger $j_*$ (F83, RF12, FR13) than ellipticals, and in the $j_*$-$M_*$ plane, in the mass range $10^{8.4} <M_* < 10^{12} M_{\odot}$, {pure discs and spheroids distribute in parallel bands.}
The same happens for intermediate morphological types (see Fig. 2 of RF12).

{
We applied the RF12 estimator, described in Section 2, to determine the $j_*-M_*$ relation of the  
McConnachie's galaxies. In the case of the RF12 and CO14 samples, we used the estimates obtained in those papers,
which were based on the same method discussed in Section 2. So, we obtained the distribution of the galaxies of the quoted samples in the $j_*-M_*$ plane, in the mass range $ 10^3 \lesssim M_* \lesssim 10^{12} M_{\odot}$.
}

%
%
%
%
%

%
%

Before comparing the $\alpha$-$M_*$ relation in the ODFBC model and in the DC14 simulation, we made two needed corrections to their results. 
In fact, in the conversion $M_*$ to $V_{\rm rot}$, DC14 a) they did not take into account a factor $h^{-2}$. 
b) They used the TFR at velocities (masses) where dSphs galaxies dominate, instead of the FJ relation. 

%
%

We then studied how the inner slope changes with rotational velocity (mass), finding that in the ODFBC model the density profile inner slope steepens in systems with lower baryonic content\footnote{In the paper, with baryons we indicate gas and stars.}, similarly to the SNFF model (and DC14 simulations), and moreover with decreasing AM.
%

The DC14 value of {$\alpha$ flattens from $\alpha \simeq -1$ at $V_{\rm rot} \simeq 250$ km/s to $\alpha \simeq 0$ at $\simeq 40$ km/s,
and then steepens to $\alpha \simeq -1.2$. 
}

{The comparison of the ODFBC model predictions with those of DC14 simulations was done in two different ways. In the first approach, we calculated the change of the slope in terms of the stellar mass, and rotational velocity, for $M_*> 10^6 M_{\odot}$, using an average AM. }

{In this case the ODFBC model has a similar behavior to DC14 simulations}: $\alpha$ flattens monotonically till $\simeq 25$ km/s and then steepens. The steepening depends on a) the rate of change of AM from the rotationally supported region of the $\alpha$-$M_*$ diagram, to that where non-rotationally supported galaxies dominate; b) the decrease of baryonic content moving towards dwarf galaxies.  

{In the second approach, we calculated the $\alpha$-$M_*$, and $\alpha$-$V_{\rm rot}$ in the case of discs and spheroids. 
In the case of discs, the slope is steeper than DC14 simulations till masses $\simeq 50$ km/s, and then it is flatter. A similar behavior is shown by spheroids, when comparing with DC14 simulations. However spheroids show a slope steeper than the discs. 
}


The main difference between {our model and DC14 simulations } is the different behavior of the slope for small velocities. 
While DC14, and similarly Madau, Shen, and Governato (2014), Oh et al. (2015), predict cuspy profiles, $\alpha \simeq -1$ for masses slightly smaller than $M_* \simeq 10^6 M_{\odot}$ (cyan line), in the ODFBC model the profiles are still cored, in agreement with $\tilde{O}$norbe et al. (2015).
Then finding cores in $M \sim 10^6 M_{\odot}$ dwarves does not imply the problems for the $\Lambda$CDM model predicted by some of the indicated SNFF models.

\subsection{Discussion}

The first discussions on the matter content in our neighborhood goes back to \''{O}pik (1915), who studied, to this aim, the vertical motions of stars near the Galactic plane\footnote{He found that the sum of stars and gas explain the stars vertical motion: there was no need to further, invisible, matter.
}.
From that time on, a century passed, and the evidences that a large part of matter in the universe consists of DM.
Nevertheless the mounting evidences, DM has not been detected to date both with direct or indirect experiments (Del Popolo 2014a).  

Dwarf galaxies, satellites of the MW, have showed to be very important in the indirect search of DM, because they are relatively close to us, and because they are dominated by DM. 

At the same time, the study of these objects has shown discrepancies among the predictions of the simulations of the $\Lambda$CDM model and observations (cusp-core problem; missing satellite problem; too-big-to-fail problem, etc). 

Recently, Pe$\tilde n$arrubia et al. (2012) studied the energetics of the cusp to core transformation in the SNFF model. The quoted transformation requires $10^{53}-10^{55}$ ergs. This imply the need of thousands of SNs to convert cusps into cores, and this disagrees with the dSphs low efficiency of star formation. Moreover, the demand of a large number of SN, i.e. a high star formation efficiency (SFE) to solve the cusp-core problem, is in tension with the demand of a low SFE to solve the TBTF problem. The tension previously indicated would worsen, if fainter dwarves would be cored. Obviously, this issue is raised in the SNFF scenario, in which SNs feedback is responsible of the cusp erasing, but not in the ODFBC. 
At the same time, SPH simulations of the cusp to core transformation (e.g., Governato et al. 2012) show that dwarves with $M_*< 10^6 M_{\odot}$ must be cuspy (e.g., Madau et al. 2014). Then finding a cored profile in dwarves of that mass would imply serious problems for the $\Lambda$CDM model, in the framework of the SNFF scenario.     

This issue shows the importance of determining the inner structure of dwarves in understanding the nature of DM. 
As we discussed in the introduction, the dSphs structure is still an open debate due to the difficulty to distinguish cusps from cores (e.g., Strigari et al. 2014). de Bruijne (2012) indicated GAIA as a possible solutions to the riddle, and Takada (2010) the
Subaru Hyper-Supreme-Camera. However, both instruments can solve the quoted problem only for larger dwarves, like Sagittarius (Richardson et al. 2014). The study of density profiles is often based on Jeans equations, which is subject to a degeneracy between the anisotropy parameter and the density profile.
This parameter cannot be determined by means of line of sight components of the stars velocity, and the 2D projection of stars radius. Better results could be obtained measuring one of the three velocity coordinates, and two of the three position coordinates
(see Battaglia et al. 2013). The density slope at half-light radius could be obtained having information on the proper motions of the dwarves stars (Strigari et al. 2007), but this is challenging even with GAIA (Richardson et al. 2014).

Despite the technical problems in having a clear cut knowledge of the inner structure of dwarves, this issue remains of fundamental importance. As we have shown in the present paper, if cores are formed through the ODFBC mechanism, the transformation of the cusp to the core starts before gas forms stars, and is more efficient. This means 
that finding cored dwarves with $M_*$ slightly smaller than $10^6 M_{\odot}$ will not result in serious problems for the $\Lambda$CDM model, and that to show disagreement between the $\Lambda$CDM model and observations one has to find cores in dwarves with smaller masses than those indicated by the SNFF mechanism in the simulations of DC14. This last point is in agreement with $\tilde{O}$norbe et al. (2015). Their simulations of dwarf galaxies of $10^4-10^6 M_{\odot}$ shows that the density profile is cuspy in a simulated ultra faint dwarf of $2 \times 10^4 M_{\odot}$.

%
%
\newpage


\section*{Appendix A: Model}\label{sect:app}

The model used to determine the $\alpha$-$M_*$ ($\alpha$-$V_{\rm rot}$) relation was introduced in DP09, and Del Popolo \& Kroupa (2009) 
%
and then applied in several other papers to study the universality of the density profiles (Del Popolo 2010, 2011), the density profiles in galaxies (Del Popolo 2012a, Del Popolo \& Hiotelis 2014) and clusters (Del Popolo 2012b, 2014b), and the inner surface-density of galaxies (Del Popolo, Cardone \& Belvedere 2013). 
The model is a semi-analytical model including an improved secondary infall model (SIM) (e.g., Gunn \& Gott 1972; Hoffman \& Shaham 1985; Del Popolo\& Gambera 1997;  Williams et al. 2004; Ascasibar et al. 2004; Hiotelis \& Del Popolo 2006,2013; Cardone
et al. 2011; Del Popolo et al. 2013a,b,c). This last takes into account random and ordered AM effects (Ryden \& Gunn 1987; Ryden 1988), exchange of energy and AM from baryons to DM through DF  (e.g., El-Zant et al. 2001, 2004), 
adiabatic contraction of DM (Blumenthal et al. 1986; Gnedin et al. 2004; Gustafsson et al. 2006). The model was further improved to take into account cooling, star formation, supernovae feedback, and reionization (e.g., Del Popolo \& Hiotelis 2014) (see the following). 

{
The model can be summarized as follows. 
Initially the proto-structure is in the linear phase, containing DM and diffuse gas. The proto-structure expands to a maximum radius then re-collapses, first in the DM component that forms the potential well in which baryons will fall. Baryons subject to radiative processes form clumps, which collapse to the centre of the halo while forming stars. In the collapse phase baryons
are compressed (adiabatic contraction) (at $z \simeq 5$ in the case of a $10^9 M_{\odot}$ galaxy (DP09)) so making the DM profile more cuspy. The clumps collapse to the galactic centre, because of dynamical friction (DF) between baryons and DM, transferring energy and AM to the DM component (El-Zant et al. 2001, 2004), increasing its random motion  and producing a predominant motion of DM particles outwards, reducing the central density\footnote{Stars are also subject to the same process (Read \& Gilmore 2005; Pontzen \& Governato 2012; Teyssier et al. 2013).}. The cusp is heated, and a core forms. 
This is the main mechanism of core formation in dwarf spheroidals and spirals\footnote{In the case of giant galaxies, as discussed, the inner profiles are steeper due to the deepening of potential well.}. 

In the case of spiral galaxies the effect of the previous mechanism is amplified by the larger AM acquired by the proto-structure through tidal torques (ordered AM), and by random AM, with respect to spheroids. Supernovae feedback acts similarly to the mechanism now discussed (see Pontzen \& Governato 2014) contributing to the core formation (see Madau, Shen \& Governato 2014).

In the following stage, supernovae explosions produce expulsion of gas in different events, clearly connected to the explosion of each supernova, leaving a lower stellar density with respect to the beginning. Feedback destroys the smallest clumps soon after a small part of their mass is transformed into stars (see Nipoti \& Binney 2015).


}

\section*{A1: Density profile formation}\label{sect:density}

{
The model follows the evolution of a perturbation starting from the linear phase, expanding with the Hubble flow, from 
initial radius $x_{\rm i}$, to the phase of maximum expansion $x_{\rm m}$ (named apapsis or turn-around radius $x_{\rm ta}$) (Peebles 1980).

The final density profile is given by (e.g., Fillmore \& Goldreich 1984; Hiotelis 2002)
\begin{equation}
 \rho(x)=\frac{\rho_{\rm ta}(x_{\rm m})}{f^3} \left[1+\frac{d \ln f}{d \ln g} \right]^{-1}\;.
 \label{eq:dturnnn}
\end{equation}
where $f(x_{\rm i})$ is the collapse factor of a shell, namely the ratio between the final radius and the turn-around radius, and  
as shown in DP09, depends on 
the radial velocity of the shell.
This radial velocity can be obtained by integrating the equation of motion of the shell:
\begin{equation}
\frac{dv_r}{dt}=\frac{h^2(r,\nu)+j_{\rm r}^2(r,\nu)}{r^3}-G(r)-\mu\frac{dr}{dt}+\frac{\Lambda}{3}r\;,
\label{eq:colll}
\end{equation}
where the acceleration $G(r)= GM_{\rm T}/r^2$, $\Lambda$ the cosmological constant, $\mu$ the 
coefficient of dynamical friction, $h(r,\nu )$\footnote{We recall that $\nu=\delta(0)/\sigma$, 
where $\sigma$ is the mass variance filtered on a scale $R_f$ 
(see DP09 or Del Popolo 1996)} 
is the ordered SAM generated by tidal torques (Hoyle 1949; Peebles 1969; White 1984; Ryden 1988; Eisenstein \& Loeb 1995; Catelan \& Theuns 1996), and $j_r(r,\nu)$ the random AM (see Ryden \& Gunn 1987 and the following of the present paper).  

The ``ordered angular momentum" is calculated by obtaining the root-mean-square (rms) torque, $\tau(r)$, on a mass shell and then calculating the total SAM, $h(r, \nu)$, acquired during expansion by integrating the torque over time (Ryden 1988, equation 35) (see Sect. C2 of DP09). As shown in DP09, in the case of a galaxy like the Milky Way, the AM obtained is $2.5 \times 10^{74} g cm^2/s$ (Appendix C1, DP09).

``Random angular momentum" (see RG87; DP09, appendix C2) is taken into account assigning a SAM at turnaround (see Appendix C2 of DP09 for details).   

Eq. (\ref{eq:colll}) shows that the larger is the content of AM of a system, the larger is the collapse time of particles at a distance $r$\footnote{We recall that there is a strong correlation between the SAM of stars $j_*$, and that of dark matter $h$\footnote{$j_r$ is strongly correlated to $h$ (Ryden 1988; DP09).}, as shown by Zavala et al. (2015) (see also FR12, and Appendix C of the present paper). }.

{Angular  momentum  sets  the  shape  of  the  density  profile in the inner regions. For a given mass, reducing the content of AM of the halo, the profile becomes steeper at the center. The steepening is easily explained in terms of an AM barrier. The halo's central density is built up by shells having pericenters close to the halo center. All particles having large angular momenta are prevented from coming close to the halos center. As a consequence they do not contribute to the central density, and this has the effect of flattening the density profile. Conversely, steeper central cusps are produced by more material having low AM. 

This simple process has been discussed in several analytical papers (Nusser 2001; Hiotelis 2002; Le Delliou \& Henriksen 2003; Ascasibar et al. 2004; Williams et al. 2004; DP09), and simulations (e.g., Polisenky \& Ricotti 2015). 

%
%

At the same time, higher peaks (larger $\nu$), which are progenitors of more massive halos, have greater density contrast (deeper potential) at their center, and so the shells do not expand far before beginning to collapse. This reduces the AM, with respect to the standard SIM (Gunn \& Fott 1972) and allows halos to become more concentrated (see section 3 of DP09, and section 5 of Williams et al. 2004 for a wider discussion). 

In summary, the shape of the halo density profile depends on its mass, and AM (also on baryon fraction, and dynamical friction). The relative role of mass, and AM, is shown in Fig. 1 of DP09, that of mass, AM, and baryon fraction in Fig. 1, 2 of Del Popolo (2012a), and Fig. 1 of Del Popolo (2012b) (see also Fig. 1 of Williams et al. 2004), 
while that of mass, AM, and dynamical friction in Fig. 1, and 2 of Del Popolo et al. (2014).

As a consequence, a dwarf spiral, despite having less AM than a giant spiral, has a flatter profile, because it has a shallower potential than a giant spiral. As discussed, in the formation phase, shells constituting the object can move far away from the center, acquire more AM, and as a consequence the DM particle tend to remain on larger orbits giving rise to flatter profiles. The same cannot happen in giant spirals, having a deeper potential well.
 
This is similar to what happens in the supernovae feedback mechanism. In dwarf galaxies, the profile is flatter with respect to giant galaxies, even if the last have a larger content of stars. This because in dwarf galaxies the expanding process of DM produced by SN explosions is not stopped or slowed down by the deep potential wells typical of giant galaxies.

The previous arguments explain why disc galaxies have shallower inner profiles than spheroids. In fact, for a given mass, $M$, the inner slope depends on the AM of the galaxy: the larger it is, the flatter is the profile. Since spheroids have on average less AM than discs, this explains the difference in the inner density profile of the two classes of objects. 

}

~\\

{ 

In Eq. (\ref{eq:colll}) appears dynamical friction, which is calculated as described in Appendix D of DP09, and Antonuccio-Delogu \& Colafrancesco (1994).  As shown in DP09, the dynamical role of dynamical friction is similar to that of AM: the larger it is, the farther DM particles remain from the center of the structure. In the case of a peak with $\nu = 3$, Fig. 11 in DP09, shows that the effect of DF increases the collapse time of $\simeq 5\%$, with respect to the standard SIM (Gunn \& Fott 1972), 
and the joint effect of DF and AM produces an increase of $\simeq 15\%$. Moreover, dynamical friction transfer energy of incoming clumps to DM.
~\\

When baryons cool dissipatively and collapse to the proto-structure centre, the DM is compressed, with the result of a 
steepening of the DM density profile (Blumenthal et al. 1986; Gnedin et al. 2004; Gustafsson et al. 2006). The process is dubbed adiabatic contraction (AC), and counteract the flattening produced by AM and dynamical friction. 

In the present paper, the model used to evaluate AC is that of Blumenthal, improved following Gnedin et al. (2004), who showed that numerical simulation results are better reproduced if one assumes the conservation of the product of the mass inside the orbit-averaged radius, and the radius itself. 

As usual, it is assumed that the initial density profile of DM and baryons is the same (Mo, Mao \& White 1998; Keeton 2001; Treu \& Koopmans 2002; Cardone \& Sereno 2005) (i.e., a NFW profile), and the final distribution of baryons is assumed to be a disc (for spiral galaxies) (Blumenthal et al. 1986; Flores et al. 1993; Mo, Mao \& White 1998; Klypin, Zhao \& Somerville 2002;  Cardone \& Sereno 2005), 
the model by Klypin, Zhao \& Somerville (2002) (see their subsection 2.1) for large spirals, and 
Hernquist model (Rix et al. 1997; Keeton 2001; Treu \& Koopmans 2002) (see Appendix E of DP09 for details), for spheroidal, and elliptical galaxies.


\section*{A2: Baryons, discs, and clumps}

Our model contains DM and baryons (initially) 
in the gas phase. The initial baryon fraction is set equal to the ``universal baryon fraction" 
$f_{\rm ub}=0.17\pm 0.01$ (Komatsu et al. 2009) (0.167 in Komatsu et al. 2011)\footnote {The baryonic fraction is obtained from the star formation processes described in the following section. }

In the case of spiral galaxies, the infalling gas will settle into a disc, which is initially stable.

Concerning the disc sizes and masses obtained in our model, we have to recall that we have already discussed this 
issue in Section 3.2 of Del Popolo (2014), when we discussed how the angular momentum catastrophe (AMC) was solved 
in our model (see also Fig. 3, and 4 in the quoted paper). This produces disc (size and mass) similar to those of real galaxies. 

When the disc density increases, it is prone to instabilities. As well known, a stationary gas is unstable if the 
Jeans criterion is satisfied, while in the case of a rotating disc, the shear force can provide an additional 
stabilizing force. 
%
%
{
The condition for the disc instability, and that for the formation of an in-situ clump is
\begin{equation}
Q \simeq \sigma \Omega/(\pi G \Sigma)=\frac{c_s \kappa}{\pi G \Sigma}<1\;,
\end{equation}
Toomre (1964), being $\sigma$ the 1-D velocity dispersion\footnote{$\sigma$ in most of the galaxies hosting clumps is $\simeq 20-80$ km/s}, $\Omega$ the angular velocity, $\Sigma$ the surface density, $c_s$ the adiabatic sound speed, and $\kappa$ the epicyclic frequency. 
When $Q<1$ the fastest growing mode is obtained knowing the dispersion relation for a perturbation (see Binney \& Tremaine 1987, or Nipoti \& Binney 2015, Eq. 6) from the solution of $d \omega^2/d k=0$, giving $k_{\rm inst}=\frac{\pi G \Sigma}{c_s^2}$
}

The previous condition was applied to the discs of our galaxies. 
The clumps radii obtained are given by
\begin{equation}
 R \simeq 7 G \Sigma/\Omega^2 \simeq 1 {\rm kpc}\;.
\end{equation}
(see Krumholz \& Dekel 2010).

For marginally unstable discs ($Q \simeq 1$), for which the velocity dispersion is maximal, and the total mass is 
three times larger than the disc cold fraction, the clumps formed have a mass $\simeq 10$ \% $M_d$ (Dekel, Sari \& Ceverino 2009).

Haloes of $5 \times 10^{11} M_{\odot}$, at $z \simeq 2$ have discs of few $\times 10^{10} M_{\odot}$, which remain for 
$\simeq 1$ Gyr in a marginally unstable state, forming clumps, of masses $\simeq 10^{9} M_{\odot}$.

In the case of dwarf galaxies, clumps have masses of $\simeq 10^5 M_{\odot}$.

The clumps characteristics (density, rotation velocity) are similar to those found by Ceverino et al. (2012), (see their 
fig. 15, 16).


Energy and AM transfer from clumps to DM can flatten the profile, and
the process is the more efficient the earlier it happened, when halos were smaller. The effectiveness
of the process has been confirmed by several authors (Ma \& Boylan-Kolchin 2004; Nipoti et al. 2004; Romano-Diaz et al. 2008, 2009; DP09 Cole et al. 2011; Inoue \& Saitoh 2011; Del Popolo et al. 2014; Nipoti \& Binney 2015).

%
%

\section*{A 2.1: Clumps life-time}
{

An important issue is the clumps life-time. Clumps must live long enough to transform the cusp into a core.

Observations of galaxies at high redshift show the presence of clumpy structures, dubbed chain galaxies and 
clump clusters (e.g., Elmegreen, Elmegreen \& Hirst 2004; Elmegreen et al. 2009; Genzel et al. 2011). 
Also HST Ultra Deep Field observations found massive star-forming clumps (Guo et al. 2012; Wuyts et al. 2013) present in a large 
number of star-forming galaxies at $z=1-3$ (Guo et al. 2015), and even in other deep fields up to $z \simeq 6$ 
(Elmegreen, Elmegreen, Ravindranath 2007).

This clumpy structures should be formed from the instability of accreating gas (e.g., Noguchi 1998; Noguchi 1999; 
Aumer et al. 2010; 
Ceverino, Dekel,\& Bournaud 2010; Ceverino et al. 2012), instability connected to the presence of a very gas-reach disc. 
Radiative cooling enhanced in the dense gas induces self-gravity instability and results in the clump 
formation.

%
%

According to the hydro-dynamical simulations of Ceverino, Dekel \& Bournaud (2010), the clumps are in Jeans equilibrium, and 
rotationally supported. This leads them to conclude that they are probably long lived ($\simeq 2 \times 10^8$ Myr). 
Krumholz \& Dekel (2010) showed that if the gas is converted in stars at a rate of a few percent, as in local systems 
forming stars, in agreement with the Kennicutt-Schmidt law, the clumps are not destroyed. 
The gas is retained and converted into stars, remaining bound and so the clump migrates to the galaxy centre. 
%
%
Simulations by Elmegreen, Bournaud \& Elmegreen (2008)  reach the same conclusions of Ceverino, Dekel \& Bournaud (2010) and Krumholz \& Dekel (2010). 
%

Last but not least, an important evidence in favor of the long lived clumps comes from observations of clumps at low 
redshift, having close similitudes (radius, mass) to that observed in the high-redshift universe Elmegreen et al. (2013), and Garland et al. (2015) (see also Mandelker et al. 2015). 

{
The previous results are very robust to changes in the lifetime, or in $\sigma$. Concerning the life-time, changes till an oder of magnitude in the life-time, will not change considerably the results. As shown by Nipoti \& Binney (2015), the flattening process happens on the dynamical friction time-scale, $1.4 t_{\rm cross}/\ln \Lambda$, where the crossing time for a system with 
$M_{\rm gas} \simeq 10^7 M_{\odot}$, is $t_ {\rm cross} \simeq 21$ Myr, and the Coulomb logarithm is $\simeq 2$, implying that the life time is an order of magnitude larger than the flattening time.

Concerning $\sigma$, as previously reported values in the range 20-80 (see footnote 23), satisfy the instability condition and clumps formation.

}

}

}
}

\section*{A3: Star formation and feedback}

Gas cooling, star formation, reionization and supernovae feedback were included as done by De Lucia \& Helmi (2008) and Li et al. (2010) (Sect.~2.2.2 and~2.2.3).

Reionization, 
produces a reduction of the baryon content, whose fraction changes as
\begin{equation}
f_{\rm b, halo}(z,M_{\rm vir})=\frac{f_{\rm b}}{[1+0.26 M_{\rm F}(z)/M_{\rm vir}]^3}\;,
\end{equation}
(Li et al. 2010), where the reionization redshift is in the range 11.5-15, $M_{\rm vir}$ is the virial mass, and $M_{\rm F}$, is the ``filtering mass" (see Kravtsov, Gnedin \& Klypin 2004).
Gas cooling is processed as a cooling flow (e.g., White \& Frenk 1991) (see Sect. 2.2.2 of Li et al. 2010). 


Concerning star formation, gas settles in a disk, and 
the star formation rate is
\begin{equation}
\psi=0.03 M_{\rm sf}/t_{\rm dyn}\;,
\end{equation}
which will give rise to an 
amount of stars 
\begin{equation}
 \Delta M_{\ast}=\psi\Delta t\;,
\end{equation}
where $M_{\rm sf}$ is the gas mass above a given density threshold, which is fixed as in DC14 as $n>9.3/{\rm cm^3}$.
$\Delta t$ indicates the time-step, and $t_{\rm dyn}$ is the disc dynamical time (see De Lucia \& Helmi
(2008) for more details).

Supernovae feedback is obtained as in Croton et al. (2006).
When a SN explodes injects in the ISM a quantity of energy given by
\begin{equation}
 \Delta E_{\rm SN}=0.5\epsilon_{\rm halo}\Delta M_{\ast} \eta_{\rm SN}E_{\rm SN}\;, 
\end{equation}
where the number of supernovae per solar mass is $\eta_{\rm SN}=8\times 10^{-3}/M_{\odot}$, in the case of Chabrier IMF (Chabrier 2003), and 
$E_{\rm SN}=10^{51}$ erg is the typical energy released in a SN explosion.

After energy injection, the gas reheating is proportional to the stars formed
\begin{equation}
 \Delta M_{\rm reheat} = 3.5 \Delta M_{\ast}\;.
\end{equation}

A quantity of hot gas equal to
\begin{equation}
 \Delta M_{\rm eject}=\frac{\Delta E_{\rm SN}-\Delta E_{\rm hot}}{0.5 V^2_{\rm vir}}\;,
\end{equation}
is ejected by the halo, if $\Delta E_{\rm SN}>\Delta E_{\rm hot}$, 
being
\begin{equation}
\Delta E_{\rm hot}= 0.5\Delta M_{\rm reheat} \eta_{\rm SN}E_{\rm SN},
\end{equation}
the thermal energy change produced by the reheated gas.


%
%


We should stress that a fundamental difference between the DC14 SNFF model and our model is that the cusp flattening starts before star formation, 
and the energy source is gravitational. Stellar and supernovae feedback start when the core is in place, and those feedback processes act to disrupt the gas clouds that formed the core (similarly to Nipoti \& Binney 2015). By converse the energy source in the SNFF processes is supernovae feedback, and the flattening process starts after stars form and explode. 

%
%

Before concluding, I want to recall that we showed the robustness of the previous model in several ways: 

a)
In DP12a, its results were compared and found to be in agreement with the density profile of galaxies of $M \simeq 10^{10} M_{\odot}$
obtained in Governato et al. (2010) SPH simulations, and in Del Popolo \& Hiotelis (2014) with the SPH simulations of Inoue \& Saitoh (2011). \\

b) 
The results of the model are in agreement with several other studies of the DFBC mechanism (El-Zant et al. 2001, 2004; Romano-D´iaz et al. 2008;Cole et al. 2011; Inoue \& Saitoh 2011; Nipoti \& Binney 2015).\\

c) 
In DP12b, the model predicted correlations and shapes of the density profiles of clusters then found by Newman et al. (2013a,b). In DP10, and DP12b it was shown that the inner slope depends on mass, in agreement with SPH simulations (e.g., DC14).


\section*{A4: Dynamics of the satellites}

{
As previously discussed, the flattening of galaxies is studied through the model described in section A1 (based on DP09, Del Popolo \& Hiotelis 2014). In this model galaxies are interacting with neighbors only through tidal interaction. 

In order to take into account the environmental effects to which satellites are subject during the infall in the host halo, together with the cusp-core transformation obtained with the model described in the previous sections, we follow Del Popolo \& Le Delliou (2014). In the quoted paper, some of the small scale problems of the $\Lambda$CDM model were studied combining the model of cusp-core transformation, described in the section A1, (based on DP09, and Del Popolo \& Hiotelis 2014) with a semi-analytic model that follows the substructure evolution within DM halos. It takes into account the effects of DF, tidal stripping and tidal heating on the infalling satellites. The model is basically the Taylor \& Babul (2001) (TB01) model with small changes coming from a similar model by Penarrubia et al. (2010) (P10). 

The model is described in Del Popolo \& Le Delliou (2014). Pullen, Benson \& Moustakas (2014) used a very similar model to study the evolution of dark matter only subhaloes. That model is part of the code GALACTICUS (Benson 2012), a semi-analytic code of galaxy formation.

}

\section*{A5: Formation of spiral and spheroids}

{

Until recently, in the traditional picture, astrophysicists thought that the major morphological features of a galaxy are determined by the assembly history and net spin of its surrounding dark matter halo: whereas discs would form within quiescently built halos whose AM content exceeds the average of the whole population, mergers would create early type elliptical galaxies. 

%
%


Sales et al. (2012) cosmological hydro-dynamical simulations, reached different conclusions.
{
According to the quoted authors, discs form when the infalling gas flows in with similar AM to that of the material that accreted earlier,  while spheroids form when the spin of the newly-accreted gas is misaligned with that of the galaxy.
}
%
%
Galactic morphology is due to the interplay between the tidal field and the material which gives rise to the galaxy.


In our model, spheroids form as in the primordial monolithic collapse from the transformation of individual gas clouds of non-rotating gas into stars (top-down model) (Eggen et al. 1962 (ELS); Gunn \& Gott 1972; Arimoto \& Yoshii 1987) and variations thereof (Peacock et al. 1998; Jiminez et al. 1998).

Discs originate as in the monolithic collapse from the dissipational collapse of clouds with scant star formation (Larson 1976)
endowed with high AM which is conserved in the collapse phase (Fall \& Efstathiou 1980; Mo, Ma, \& White 1998). 
As discussed in section A1, the final baryon configuration is fixed in the adiabatic contraction phase
to be a disk, or a spheroid.  
The previous basic model, was improved following Sales et al. (2015) results. 
{
As in Sales et al. (2012), we measure morphology in terms of 
\begin{equation}
\kappa_{\rm rot}
=\frac{K_{\rm rot}}{K}=\frac{1}{K} \sum \frac{m}{2} \left( \frac{j_z}{R} \right)^2
\end{equation}
where $K_{\rm rot}$ is the kinetic energy in ordered motions, $j_z$ the component of AM orthogonal to the disk, $m$ the mass of a fraction of the system (e.g., the mass contained in a mass shell), and $R$ its radius.
Discs with perfectly circular motions have $\kappa_{\rm rot} \simeq 1$, and spheroids have $\kappa_{\rm rot}\ll1$. As in Sales et al. (2012), disk dominated galaxies have $\kappa_{\rm rot}>0.7$, and spheroids $\kappa_{\rm rot}<0.5$. We then define, at turn-around, $\cos{\theta}$ as the angle, {
between the AM enclosed by different mass shells, $m$, and the total one, $m_{\rm tot}$ (see Sales et al. 2012). } At turn-around, we calculate the ratio $m/m_{\rm tot}$, and $\cos{(\theta)}$ and we impose that $\cos{(\theta)}$ in terms of $m/m_{\rm tot}$ reproduces the behavior of the $\cos{(\theta)}$-$m/m_{\rm tot}$ relation given in Fig. 9 of Sales et al. (2012). We then choose, in the quoted systems, those that for a $M_*$ have $j_*$ in agreement with the $j_*$-$M_*$ relation of Fig. 1.
}

~\\
Concerning satellite dwarf galaxies, the action of tidal stripping, tidal heating, and ram-pressure stripping on a rotating late-type dwarf can remove the majority of its AM and gas transforming it in a non-rotating and quiescient early-type galaxy (Mayer et al. 2006; Schroyen et al. 2011). If tidal effects are not strong, and in absence of ram-pressure stripping, a late-type dwarf can be transformed into a dIrr/dE transition type dwarf (see the Introduction of Schroyen et al. 2011).

\section*{A6: TF, FJ, and the $M_*$-$M_ {\rm halo}$ relations}

In this section, we compare the stellar mass-TF, FJ relation obtained by Dutton et al. (2010), and the $M_*$-$M_{\rm halo}$ relation of Moster et al. (2013), and Munshi et al. (2013) with the same relations obtained with the model of the present paper. In the top left panel of Fig. 4,  the short dashed line represents the Dutton et al. (2010) TF, and the thin solid lines represents the uncertainty (0.027 on the zero point, 0.003 on the slope, and intrinsic scatter 0.05 dex on 
$V_{\rm rot}$).

The thick solid line represents the TFR obtained using the model used in the present paper. The plot shows a  good agreement of our TFR with that of Dutton et al. (2010). 

The top right panel represents the comparison of our FJR with that of Dutton et al. (2010). Symbols are like the previous plot, and again a good agreent is found. In this case, the uncertainty on the slope is 0.020, and the scatter is 0.071 dex in $\sigma(R_{50})$. 

In order to make comparisons with smaller values of the mass, in the bottom left panel, we compare our TF relation with that of McGaugh \& Wolf (2010), while in the bottom right panel we compare our FJ relation to that of C014 (see the Fig. 4 caption for details).

In Fig. 5, we plot the stellar mass vs halo mass relation with those of Moster et al. (2013), and Munshi et al. (2013) (see the Fig. 5 caption for details).


%
%

\begin{figure*}
\centering
\hspace{-2cm}
\includegraphics[width=180mm]{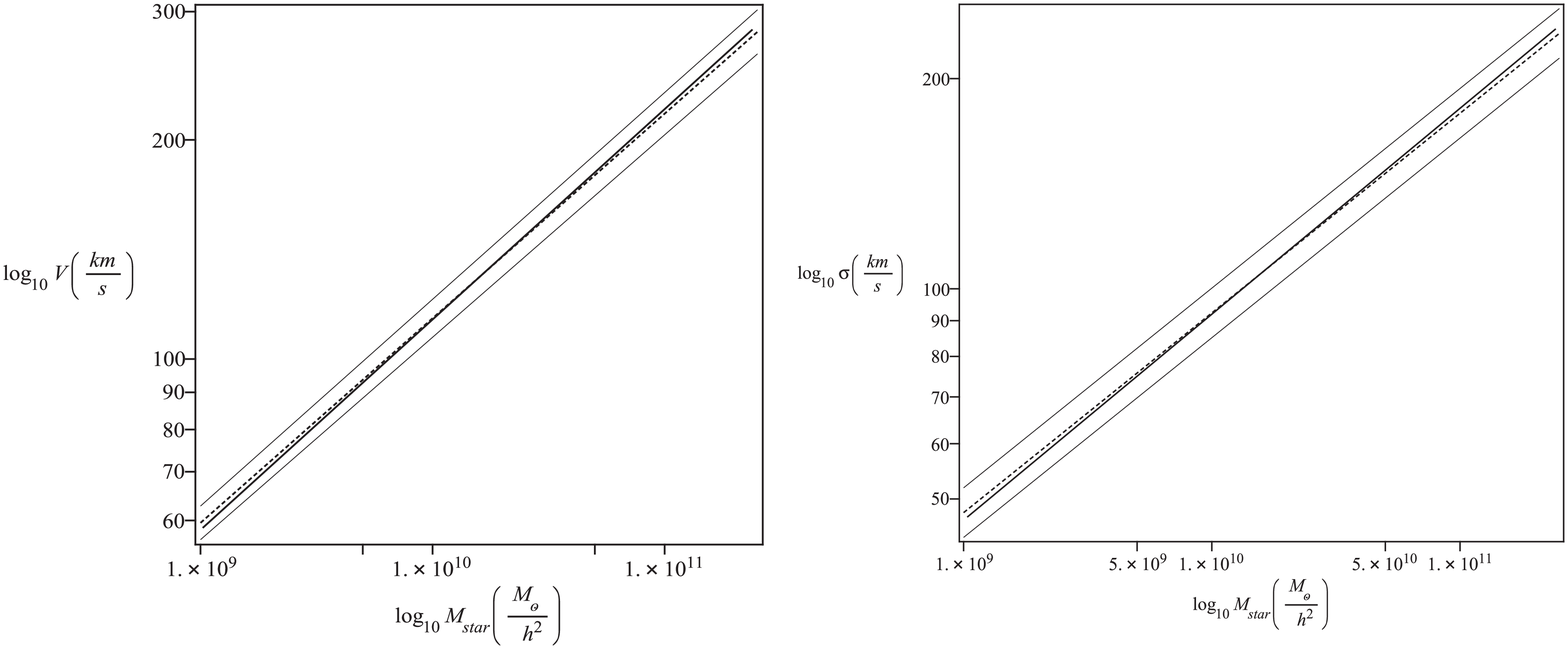} 
\includegraphics[width=169mm,height=70mm]{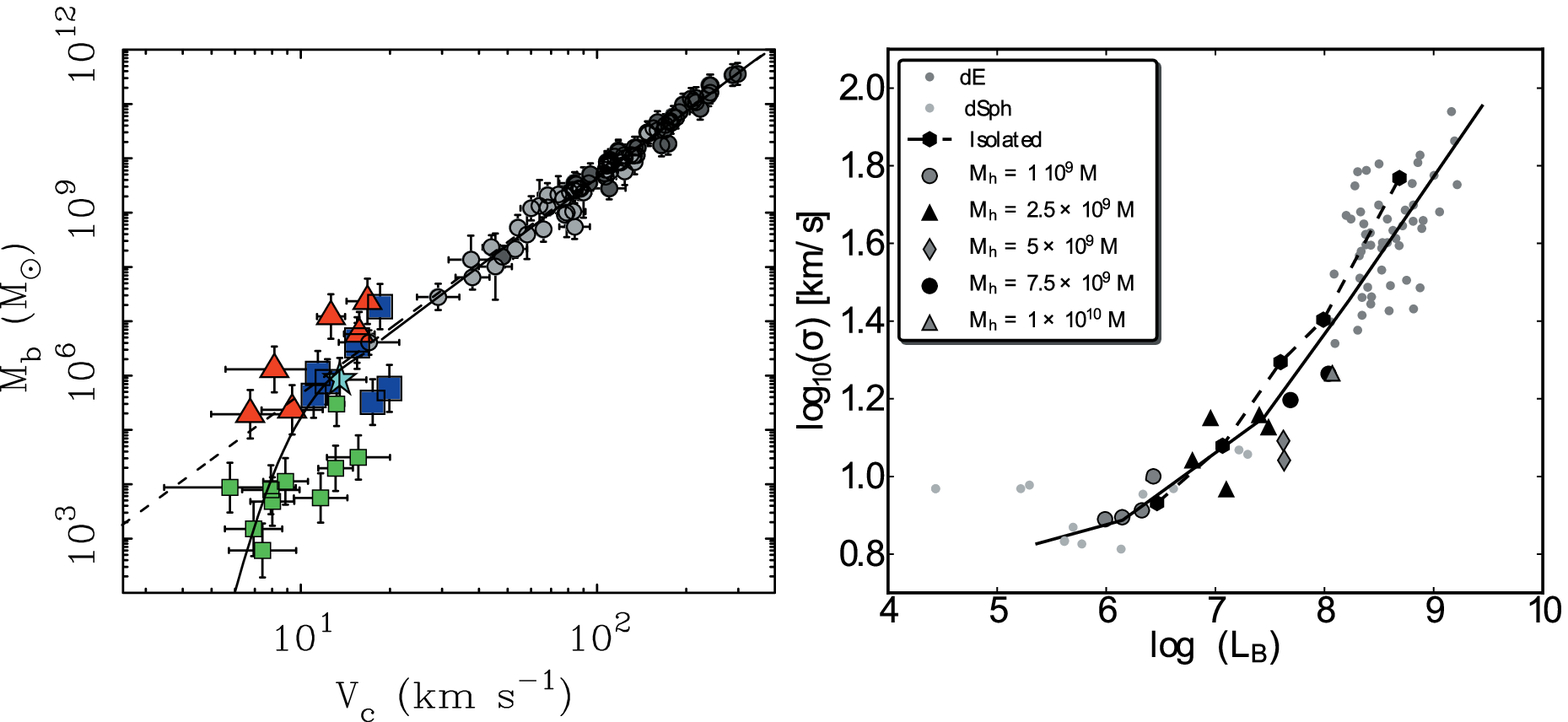}
\vspace{2cm}
\caption[]{ Stellar mass-TF, FJ relation, and $M_*$-$M_{\rm halo}$ relation. Left top panel: the short dashed line is the Dutton et al. (2010) TF, and the thin solid lines represents the uncertainty (0.05 dex in 
$V_{\rm rot}$). The thick solid line is the TF relation obtained using this paper model. The uncertainty is 0.071 dex in 
$\sigma$. Right top panel: as in the left panel but for the FJ relation. Bottom left panel: comparison of the baryonic TF relation of this paper (solid line), with that of McGaugh \& Wolf (2010). The symbols are as follows: green squares (ultrafaint dwarfs (UFD)), dark gray points (star-dominated spirals), gas-dominated disks (light gray points), large blue squares (classical dwarfs), red triangles (M31 dwarfs), and the light blue star (Leo T). The dashed line is the best fit excluding the UFD (see McGaugh \& Wolf 2010). Bottom right panel: the FJ relation of C04. Circles are observed galaxies, the other symbols are simulated galaxies. The dashed line connects the isolated galaxies, and the solid line is our FJ relation.
}
\end{figure*}

\begin{figure*}
\centering
\includegraphics[width=100mm]{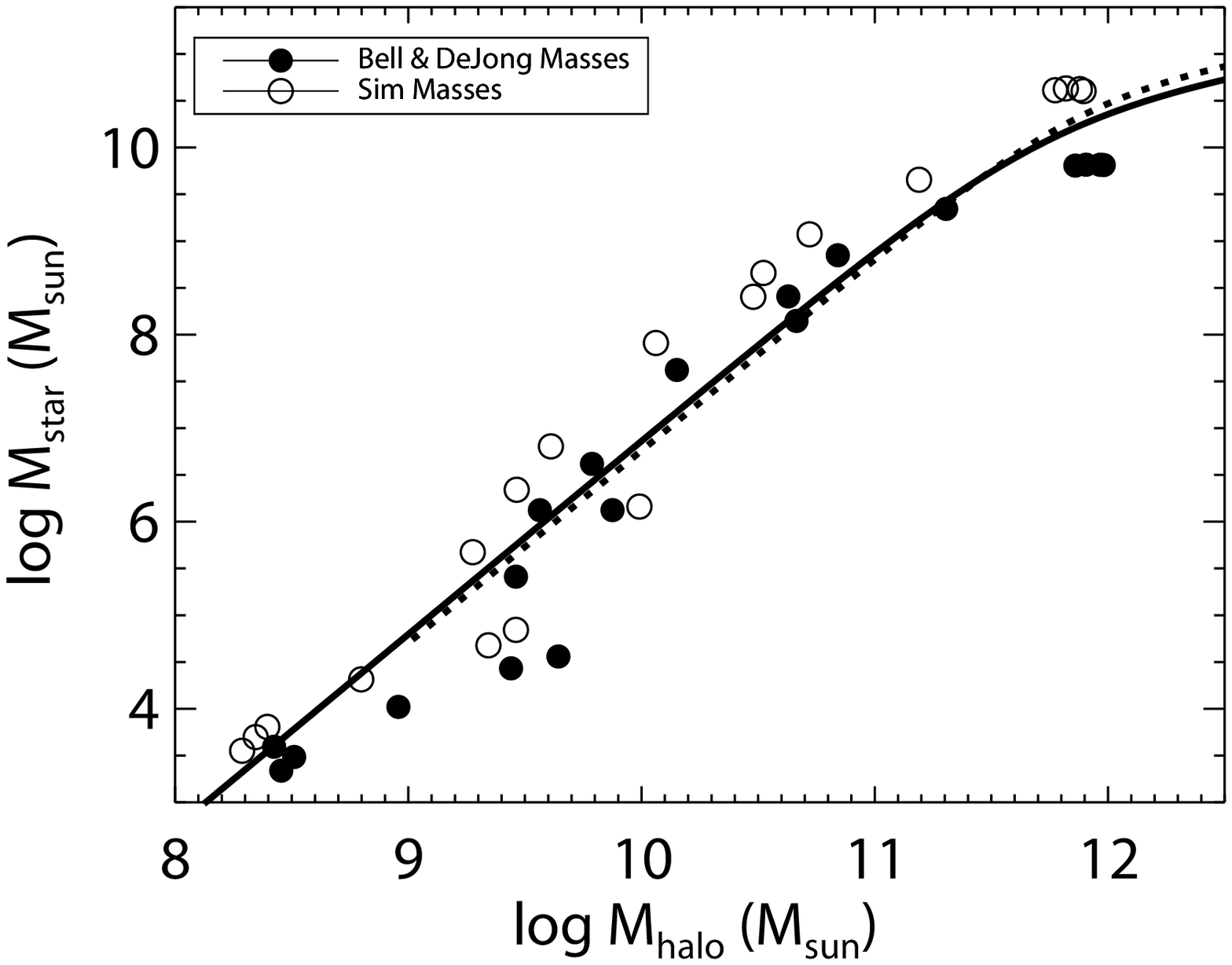} 
\caption[]{ Stellar mass vs halo mass (SMH) relation. Dashed-line (Moster et al. 2013), solid line (our SMH), open dots (SMH with stellar masses measured directly from Munshi et al. (2013) simulations), and solid dots (SMH with stellar masses measured through Petrosian magnitudes (Munshi et al. 2013)).
}
\end{figure*}

}
}

\section*{Appendix B: Galaxies in and around the Local Group}


%

Apart for the MW and Andromeda, in the McConnachie's sample there are two other spiral galaxies, namely M33, and NGC 300. dSph galaxies are predominant, followed in number by dIrr galaxies, mixed morphologies dIrr/dSph, and dE/dSph, a few Irregulars (Irrs), and one cE, namely M32. 

In this paper, as MC12, and Grebel et al. (2013), galaxies fainter than $M_V \simeq -18$ are considered dwarf galaxies. Then all the galaxies in the sample, except the MW, the LMC, NGC 55, M33, M31, and NGC 300, are dwarves. 

dIrrs are irregular, and gas-rich galaxies with $\mu_V \lesssim 23$ $\rm mag/arcsec^2$, $M_{\rm HI} \lesssim 10^9 M_{\odot}$, and 
$M_{\rm t} \lesssim 10^{10} M_{\odot}$ (Grebel 2001). Low mass dIrss do not show measurable rotation, and could evolve to become dSphs, while the more massive have often rigid body rotation, but not shear, like the differential rotation in the discs of spirals, and may go on forming stars along all their history (Hunter 1997, Grebel 2001).

dEs have spherical or elliptical shape, and are fainter than $M_V=-17$ $\rm mag$, have $\mu_V \lesssim 21$ $\rm mag/arcsec^2$, $M_{\rm HI} \lesssim 10^8 M_{\odot}$, and $M_{\rm t} \lesssim 10^{9} M_{\odot}$ (Grebel 2001).

dSphs are gas-poor, and have a low-surface-brightness, $\mu_V \gtrsim 22$ $\rm mag/arcsec^2$, $M_V \gtrsim -14$ mag, $M_{\rm HI} \lesssim 10^5 M_{\odot}$, and $M_{\rm t} \lesssim 10^{7} M_{\odot}$ (Grebel 2001). They are then the least massive and faintest galaxies known. 

dIrr/dSphs, are transition-types and have $M_{\rm HI} \lesssim 10^6 M_{\odot}$, much smaller than dIrrs ($10^7  \lesssim M_{\rm HI}\lesssim 10^9 M_{\odot}$). 

Classifying objects into objects having measurable rotation or not, is not an easy task. 
In the plane V-band luminosity-mean metallicity of red giants (see Fig. 1 of Grebel et al. 2003, and Fig. 7 of Mateo 1998), massive dIrrs, and dSphs are grouped into different areas of the plane. dIrr/dSphs are generally closely located to dSphs, and they differ from dSphs only for the gas content. This is the case of the DDO 210, Phoenix, Antlia, LGS 3, and KKR 25. Peg DIG which is a transition type galaxy, differently from the previous ones, is located in the dIrr's locus. GR 8, and Leo A are faint dIrrs, and differently from massive dIrrs are close to the dSphs locus.
It is often assumed that if the ratio between the rotational velocity and velocity dispersion, $V_{\rm rot}/\sigma$, is $\lesssim 1$, random motion should dominate on rotation (e.g., Lo, Sargent, \& Young 1993), and usually dSphs, fulfill this criterion. However, there are exceptions to it (e.g., Phoenix).

Because of the similitude among low mass dIrrs, dIrr/dSphs and dSphs, one can conclude that the more massive dIrrs are characterized by rigid body rotation, while the low mass dIrrs, and dIrr/dSphs have low rotational support (Grebel 2001; Grebel et al. 2003). Objects like GR 8, Leo A, LGS 3, Sag DIG, and DDO 210 do not show systematic rotation in HI (Lo et al. 1993; Young\& Lo 1997)\footnote{Notice that according to Mateo 1998, there are two rotating dSphs, having $V_{\rm rot}/\sigma \lesssim 1$: NGC 147 (Bender et al 1991) and UMi (Hargreaves et al 1994b, Armandroff et al 1995). The rotation of UMi could be due to streaming motions generated by tides (Oh et al 1995, Piatek \& Pryor 1995). Moreover according to Mateo 1998, Carignan et al. (1990) GR 8 
shows evidence for rotation even if $V_{\rm rot}/\sigma \lesssim 1$. Young \& Lo (1996) arrived to similar conclusions for Leo A.}.

In the case of Phoenix, $V_{\rm rot}/\sigma=1.7$, but the HI cloud has a small velocity gradient. This can be due to ejections from Phoenix (St-Germain et al. 1999), or to rotation. Peg DIG has $V_{\rm rot}/\sigma=1.7$. In the case of KKR 25 and Antlia the $V_{\rm rot}/\sigma$ is not measured, but the measurement of HI rotation indicates that $V_{\rm rot} \lesssim 5$ km/s (Grebel et al. 2003).

Summarizing the previous discussion, dIrr/dSphs have similar characteristics to dSphs, and very small rotation. Massive dIrrs are rotationally supported, while small mass dIrrs are not. 

One could then empirically ``define" the boundary among rotation and no rotation by dIrrs such as GR 8, Leo A, LGS 3, SagDIG, and DDO 210. If one adopts the dynamical masses compiled by MC12, this would imply that galaxies with dynamical masses below a few times $10^7 M_{\odot}$ are no longer rotation-dominated, whereas galaxies with masses $> 10^8 M_{\odot}$ do show measurable rotation (Grebel, private communication). {For precision's sake, we should report that the dSphs dominate in systems like M31 or the MW, while in general isolated dwarf galaxies in the local universe have different characteristic. However, we already discussed that the $j_*-M_*$ relation in the case of isolated dwarfs have similar characteristics to that obtained from the MC12 sample.}


Following the previous criteria, we divided the MC12 sample into fast rotators (FR), and slow rotators (SR)
This is shown in Tab. 1. The first column gives the name of the galaxy, the second the stellar mass, the third the dynamical mass, the fourth the specific stellar AM, 
and the fifth the morphological (Hubble) type. The galaxies 
are divided into two groups (slow rotators, and fast rotators, respectively upper and lower part of the table).

Fig. 1 plots the distribution FRs (histogram in red), and SRs (histogram in blue).  



The maximum mass of non-rotating objects is $10^{7.8} M_{\odot}$. 


This mass threshold between rotation and non-rotation dominated galaxies, should not be considered as an abrupt 
transition between the two regimes, 
but as the mass under which the AM gradually decreases from the values of discs to that of spheroids, as shown in Fig. 2 of RF12.

%
%


\section*{Appendix C: Tidal torque theory, and the $j_*-M_*$ relation}

From the tidal torque theory (TTT) (Hoyle 1949, Peebles 1969, Doroshkevich 1970, White 1984) we know that $j \propto M^{2/3}$. 
{They considered a simple model of transformations of gas into stars taken into account by the parameter $f_*$ (fraction of the initial gas that transforms in stars), and an idealized model of the baryon AM transfer, parameterized through $f_j$\footnote{For example by AM transfer from stars to DM.}. 
}

Using the two quoted parameters, the $j-M$ relation for dark matter will become the $j_*$-$ M_*$ for stars: 
\begin{equation}
j_*=2.92 \times 10^4 f_j f_*^{-2/3} \lambda \left(\frac{M_*}{10^{11} M_{\odot}}\right)^{2/3}  {\rm km/s kpc},
\label{eq:estim1}
\end{equation}
where $\lambda$ is the spin parameter, with {$\lambda=<\lambda>=0.035$. The average value of $f_j f_*^{-2/3}$ ($<f_j f_*^{-2/3}>$) can be obtained using a sample of galaxies. RF12 found $<f_j f_*^{-2/3}> \simeq 1.9$ for Sb-Sm galaxies, and $<f_j f_*^{-2/3}> \simeq 0.5$ for ellipticals.
A more precise analysis must take into account a dependency on the efficiency of star formation. This is obtained using the 
$\langle f_\star\rangle(M_\star)$ given by Dutton et al. (2010).  
So, for galaxies with $M_* > 10^{8.6} M_{\odot}$, RF12 got the following $\langle f_\star\rangle(M_\star)$, }
\begin{equation}\label{eqn:fsMltg}
\langle f_\star\rangle(M_\star) = \frac{f_0 \left(M_\star/M_0\right)^{1/2}}{\left[1+\left(M_\star/M_0\right)\right]^{1/2}},
\end{equation}
(RF12), for spiral galaxies. The constant $M_0$ is given by $\log\,(M_0/M_\odot) \simeq 10.8$, and $f_0 \simeq 0.33$.
In the case of elliptical galaxies: 
\begin{equation}\label{eqn:fsMetg}
\langle f_\star \rangle (M_\star) = \frac{f_0 \left(M_\star/M_0\right)^{0.15}}{\left[1+\left(M_\star/M_0\right)^2\right]^{1/2}},
\end{equation}
where $\log\,(M_0/M_\odot)\simeq11.2$, $f_0 \simeq 0.14$,


{Concerning $f_j$ RF12 considered two scenarios. In the first one, $f_j$ is variable, and depending on morphology, based on the fact that the different $j_*-M_*$ relation of galaxies and spirals could be connected to different retention in the SAM. 
In the second scenario,  $f_j$ is constant, based on the fact that a scatter in the spin parameter, $\lambda$, is expected. Then they obtained average values for $f_j$, which are similar in the two scenarios.
}

As previously reported, discs and ellipticals are distributed on two bands with an offset in $j_*$. 

{
RF12 and FR13 conclusions are confirmed by Teklu et al. (2015) and Zavala et al. (2015). In particular, in the second paper, the authors showed that spheroids are fundamentally formed by stars formed before turn-around, while
discs are constituted by stars forming after turnaround.
}

\begin{table*}
  \centering
\tiny
  \caption{
Parameters for McConnachie (2012) galaxies. The first column gives the name of the galaxy, the second the stellar mass, the third the dynamical mass, the fourth the specific stellar AM, 
and the fifth the morphological (Hubble) type. 
%
%
The values of $j_*$ were calculated similarly to the way RF12 calculated the SAM for their sample, as described in Section 2.
%
%
}
\begin{tabular}{rrrrrrr}
\hline 
 Name  & $\frac{M_*}{10^6 M_{\odot}}$ & $\frac{M_{\rm dyn}}{10^6 M_{\odot}}$ & $\log(j_*)$ (kpc km/s) &       Morphology &  \\
\hline

&       &       & {\bf SRs}
 &       &       &  \\

\hline

Canis Major                                & 49    &       & 0.804 &  \\
    Sagittarius dSph     & 21    & 190   & 0.3013 & dsph \\
    Segue (I)       & 0.0003 & 0.26  & -2.139 & dsph \\
    Ursa Major II        & 0.0041 & 3.9   & -1.916 & dsph \\
    Bootes II                & 0.001 & 3.3   & -2.139 & dsph \\
    Segue II                 & 0.0009 & 0.23  & -1.93 & dsph \\
    Willman 1             & 0.001 & 0.27  & -1.857 & dsph \\
    Coma Berenices    & 0.0037 & 0.94  & -1.617 & dsph \\
    Bootes III             & 0.017 &       & -1.493 & dsph \\
    Bootes (I)            & 0.029 & 0.81  & -1.11 & dsph \\
    Draco                   & 0.29  & 11    & -0.379 & dsph \\
    Ursa Minor           & 0.29  & 9.5   & -0.502 & dsph \\
    Sculptor                 & 2.3   & 14    & 0.2301 & dsph \\
    Sextans (I)              & 0.44  & 25    & -0.27 & dsph \\
    Ursa Major (I)        & 0.014 & 11    & -1.116 & dsph \\
    Carina                      & 0.38  & 6.3   & -0.369 & dsph \\
    Hercules                   & 0.037 & 2.6   & -0.893 & dsph \\
    Fornax                      & 20    & 56    & 0.7748 & dsph \\
    Leo IV                      & 0.019 & 1.3   & -1.183 & dsph \\
    Canes Venatici II      & 0.0079 & 0.91  & -1.287 & dsph \\
    Leo V                         & 0.011 & 1.1   & -1.27 & dsph \\
    Pisces II                      & 0.0086 &       & -1.62 & dsph \\
    Canes Venatici (I)       & 0.23  & 19    & -0.518 & dsph \\
    Leo II                           & 0.74  & 4.6   & -0.205 & dsph \\
    Leo I                             & 5.5   & 12    & -0.034 & dsph \\
    Andromeda IX               & 0.15  & 6.5   & -0.879 & dsph \\
    Andromeda XVII           & 0.26  &       & -0.475 & dsph \\
    Andromeda I                  & 3.9   & 44    & 0.1725 & dsph \\
    Andromeda XXVII       & 0.12  &       & -0.886 & dsph \\
    Andromeda III                & 0.83  & 6.1   & -0.025 & dsph \\
    Andromeda XXV          & 0.68  &       & -0.123 & dsph \\
    Andromeda XXVI         & 0.06  &       & -0.804 & dsph \\
    Andromeda XI                 & 0.049 & 1.9   & -0.854 & dsph \\
    Andromeda V                 & 0.39  &       & -0.415 & dsph \\
    Andromeda X                 & 0.096 & 2.3   & -0.727 & dsph \\
    Andromeda XXIII           & 1.1   &       & -0.14 & dsph \\
    Andromeda XX                & 0.029 &       & -0.957 & dsph \\
    Andromeda XII                & 0.031 & 1.2   & -1.037 & dsph \\
    Andromeda XXI               & 0.76  &       & -0.202 & dsph \\
    Andromeda XIV                & 0.2   & 6.1   & -0.539 & dsph \\
    Andromeda XV                  & 0.49  & 16    & -0.449 & dsph \\
    Andromeda XIII                 & 0.041 & 11    & -1.154 & dsph \\
    Andromeda II                      & 7.6   & 36    & 0.191 & dsph \\
    NGC 185                            & 68    & 150   & 1.0899 & dEdsph \\
    Andromeda XXIX             & 0.18  &       & -0.792 & dsph \\
    Andromeda XIX                 & 0.43  &       & -0.421 & dsph \\
    Andromeda XXIV               & 0.093 &       & -0.955 & dsph \\
    Andromeda VII                    & 9.5   & 42    & 0.2667 & dsph \\
    Andromeda XXII                 & 0.034 &       & -0.888 & dsph \\
    LGS 3                                   & 0.96  & 17    & -0.093 & dIrrdsph \\
    Andromeda VI                      & 2.8   &       & 0.0899 & dsph \\
    Andromeda XVI                   & 0.41  & 7.9   & -0.457 & dsph \\
    Andromeda XXVIII              & 0.21  &       & -0.496 & dsph? \\
    Phoenix                      & 0.77  &       & -0.306 & dIrrdsph \\
    Cetus                                       & 2.6   & 120   & 0.0897 & dsph \\
    Pegasus dIrr                    & 6.61  &       & 0.3878 & dIrrdsph \\
    Leo T                                      & 0.14  & 3.9   & -0.655 & dIrrdsph \\
    Leo A                                     & 6     & 25    & 0.0668 & dIrr \\
    Andromeda XVIII                & 0.63  &       & -0.447 & dsph \\
    Tucana                                 & 0.56  & 41    & -0.377 & dsph \\
    Sagittarius dIrr                               & 3.5   &       & 0.1123 & dIrr \\
    UGC 4879                     & 8.3   &       & 0.1276 & dIrrdsph \\
    Sextans B                    & 52    &       & 0.6088 & dIrr \\
    Antlia                       & 1.3   &       & -0.142 & dIrr \\
    HIZSS 3(A)                   &       &       &       & (d)Irr \\
    HIZSS 3B                     &       &       &       & (d)Irr \\
    KKR 25                       & 1.4   &       & 0.0317 & dIrrdsph \\
    ESO 410- G 005               & 3.5   &       & 0.3252 & dIrrdsph \\
    ESO 294- G 010               & 2.7   &       & 0.0565 & dIrrdsph \\
    KKH 98                       & 4.5   &       & 0.0238 & dIrr \\

\end{tabular}
  \label{tab:addlabel}%
\end{table*}%

\setcounter{table}{0}
\begin{table*}
  \centering
\tiny
  \caption{$(${\it Continued}$)$}
\begin{tabular}{rrrrrrr}
\hline 
 Name  & $\frac{M_*}{10^6 M_{\odot}}$ & $\frac{M_{\rm dyn}}{10^6 M_{\odot}}$ & $\log(j_*)$ (kpc km/s) &       Morphology &  \\
\hline
    UKS 2323-326             & 17    &       & 0.3555 & dIrr \\
    KKR 3                    & 0.54  &       & -0.553 & dIrr \\
    GR 8                     & 6.4   &       & 0.4571 & dIrr \\
    UGC 9128                 & 7.8   &       & 0.5618 & dIrr \\
    UGC 8508                 & 19    &       & 0.7529 & dIrr \\
    IC 3104                  & 62    &       & 0.7524 & dIrr \\
    DDO 125                  & 47    &       & 0.5286 & dIrr \\
    UGCA 86                  & 16    &       & 0.6979 & dIrr \\
    DDO 99                   & 16    &       & 0.5214 & dIrr \\
    DDO 190                  & 51    &       & 0.8028 & dIrr \\
    KKH 86                   & 0.82  &       & -0.056 & dIrr \\
    NGC 4163                 & 37    &       & 0.7644 & dIrr \\
    DDO 113                  & 2.1   &       & -0.141 & dIrr \\
    Aquarius & 1.6   &       & -0.199 & dIrrdsph \\

\hline

&       &       & {\bf FRs}
 &       &       &  \\

\hline
    LMC                                        & 1500  &       & 2.4447 & Irr \\
    SMC                                                        & 460   &       & 2.1465 & dIrr \\
    M32                              & 320   & 540   & 1.9229 & cE \\
    NGC 205                        & 330   & 420   & 1.7475 & dEdsph \\
    NGC 147                         & 62    & 93    & 1.5486 & dEdsph \\
    Triangulum                          & 2900  &       & 2.4799 & Sc \\
    IC 1613                     & 100   &       & 1.523 & dIrr \\
    NGC 6822                    & 100   &       & 1.7457 & dIrr \\
    WLM                       & 43    & 380   & 1.6526 & dIrr \\
    NGC 3109              & 76    &       & 1.7559 & dIrr \\
    Sextans A            & 44    &       & 1.4664 & dIrr \\
    NGC 55               & 2200  &       & 2.4791 & Irr \\
    NGC 300     & 2100  &       & 2.1589 & Sc \\
    IC 5152           & 270   &       & 1.6889 & dIrr \\
    IC 4662       & 190   &       & 2.0106 & dIrr \\
    IC 10            & 86    &       & 1.4036 & dIrr \\
\end{tabular}
  \label{tab:addlabel}%
\end{table*}%


\begin{acknowledgements} 
{\bf Acknowledgements} \\
The author would like to thank Morgan Le Delliou and Charles Downing for a critical reading of the paper. 
\end{acknowledgements}

\label{lastpage}

\end{document}